\documentclass[onecolumn]{article}
\usepackage{authblk}
\usepackage[a4paper, total={6in, 8in}, margin=1in]{geometry}

\usepackage[
backend=biber,
style=numeric,
sorting=none, 
maxnames=99,
minnames=99
]{biblatex}
\usepackage{graphicx} 
\usepackage{xcolor}
\usepackage{subcaption}
\usepackage{booktabs}
\usepackage{verbatim}
\usepackage[utf8]{inputenc}
\usepackage{amsmath} 
\usepackage{amssymb}

\usepackage{setspace}
\DeclareUnicodeCharacter{200B}{}

\title{Steady state, core, operational optimization of an ARC-like tokamak via plasma composition and shape}

\author[1]{A. Saltzman}
\author[1]{P. Rodriguez-Fernandez}
\author[1]{A. Ho}
\author[1]{G. Snoep}
\author[1]{J. Han}
\author[1]{J. Hall}
\author[2]{M. S. Anastopoulos Tzanis}
\author[3]{J. Hillesheim}
\author[3]{A. J. Creely}
\author[3]{P. Snyder}
\author[1]{N.T. Howard}
\affil[1]{MIT Plasma Science and Fusion Center}
\affil[2]{Oak Ridge National Lab}
\affil[3]{Commonwealth Fusion Systems}

\date{}

\begin{document}

\maketitle

\begin{abstract}
Impurity composition, plasma shape, and pedestal density all provide strong levers on fusion power. Here, we explore the ways in which their variation changes fusion power and seek to find the optimum of these parameters. The key impacts of these variables are through changes in the core turbulent transport, the density of the fuel species, the pedestal pressure, and the plasma volume. ITG stabilization due to increased amounts of impurities is observed.  The dependence of all of these parameters on the pedestal pressure is especially complicated because of the separate impacts on the peeling and ballooning modes, which can each limit the pedestal. Optimization of this multidimensional operating space is enabled by the use of Bayesian optimization, resulting in an operating point similar to ARC V3A with \textcolor{black}{$\sim$30\%} more fusion power and a higher fusion power density. Increased shaping parameters, including elongation, triangularity, and squareness are all beneficial, as is high $Z_{eff}$. \textcolor{black}{When elongation is also allowed to vary, a $\sim$65\% increase in fusion power can be achieved.} While not commonly considered, we find squareness is an important lever on fusion power. The plasma performance is limited by the Greenwald density limit constraint. This workflow developed here and demonstrated with the example of ARC V3A can readily be applied to other tokamak designs.
\end{abstract}

\section{Introduction}

ARC is a high field, tokamak power plant under development by Commonwealth Fusion Systems, the physics basis for which has recently been published \cite{hillesheimOverviewPhysicsBasis2026}. Inductive, positive triangularity operation of ARC is envisioned. As a commercial fusion power plant, achieving the desired fusion power for minimal cost is desired. Recent studies have shown the potential to improve plasma performance via modifications to impurity composition and plasma shaping \cite{rodriguez-fernandezCorePerformancePredictions2024, dunneGlobalPerformanceEnhancements2016}. Here, optimization of these parameters is performed for ARC-like plasmas with the aim of maximizing fusion power that can be achieved for a given device design, which would have a fixed cost. The physical mechanisms driving the optimal impurity composition and plasma shape are then explored. 

\subsection{Past ARC performance predictions}
 Extensive, state of the art, modeling has been undertaken to ensure fusion power targets, necessary for achieving new power production, are met \cite{howardPerformanceTransportARC2026} in ways that are compatible with the overall system design. This includes compatibility with achievable power and particle exhaust solutions \cite{eichPowerParticleExhaust2026}. The current ARC design, ARC V3A has a major radius of 4.62 meters, a minor radius of 1.18 meters, a magnetic field on axis of 11.4 T, and a plasma current of 12.0 MA. 21.5 MW of ion cyclotron resonance heating (ICRH) power is envisioned to be used during flattop. In this work, we will explore the parameter space around the ARC V3A design \textcolor{black}{with the joint goals of optimizing its as designed performance and informing future design iterations}. 

Performance modeling thus far has included low-fidelity, empirical-based 0D POPCON modeling, medium fidelity integrated modeling with the quasilinear models and high fidelity nonlinear gyrokinetic core profile predictions. The predicted fusion power from these models is between $\sim$900MW and 1300MW, with the higher fidelity models generally predicting lower performance \cite{howardPerformanceTransportARC2026}. Key sources of uncertainty in these predictions include uncertainty on the pedestal pressure boundary condition from EPED \cite{snyderDevelopmentValidationPredictive2009,snyderFirstprinciplesPredictiveModel2011} as well as the core turbulent transport \cite{howardPerformanceTransportARC2026}. 

\subsection{Importance of using physics-based transport models}
Physics-based models leverage theoretical knowledge to make predictions, which have been validated against experimental data, whereas empirical models rely on direct extrapolations from experimental data. Physics-based core turbulent transport models can make the same \cite{abbateLargedatabaseCrossverificationValidation2024}
 or more accurate \cite{angioniConfinementPropertiesLmode2022, ludaIntegratedModelingASDEX2020, staeblerAdvancesPredictionTokamak2022} predictions than empirical models. Often, empirical models used in 0D frameworks overestimate the performance of a tokamak design. This is both because critical gradients effects are not fully captured by the scaling law's power degradation term \cite{iterphysicsexpertgrouponconfinementandtransportChapter2Plasma1999,rodriguez-fernandezCorePerformancePredictions2024} and because optimistic improvements in the confinement time are assumed above the prediction of the energy confinement scaling laws. Additionally, non-physical profiles, including parabolic profiles and profiles with arbitrary temperatures are often assumed \cite{saltzmanImpactModelUncertainty2025a}; this was the case with early scoping studies for compact high-field devices \cite{sorbomARCCompactHighfield2015}.  
 
 One key shaping parameter, triangularity, is not included in most of the commonly used energy confinement time scaling laws \cite{iterphysicsexpertgrouponconfinementandtransportChapter2Plasma1999, yushmanovScalingsTokamakEnergy1990, kayeITERLmodeConfinement1997,  leeTokamakElongationHow2015}, with the exception of \cite{verdoolaegeUpdatedITPAGlobal2021}.  However, triangularity has been shown experimentally to have large impacts on performance -- both for strong positive \cite{uranoPedestalStructureHmode2014} and negative triangularities \cite{thomeOverviewResults20232024, marinoniBriefHistoryNegative2021}. Indeed, later studies of ARC-class devices, which leveraged physics-based turbulent transport modeling, showed the important predicted impact of triangularity on performance \cite{frankRadiativePulsedLmode2022, collaborationMANTANegativetriangularityNASEMcompliant2024}. Furthermore, while empirically-based 0D modeling captures the impact of dilution on fusion power, it does not capture the potentially beneficial impact of ITG stabilization \cite{migliuoloIonTemperatureGradient1992, rodriguez-fernandezCorePerformancePredictions2024}. That physics-based models can capture these effects, which empirical models do not, make them critical for use in tokamak design.

 \subsection{Operational optimization}

 There are two stages of design optimization that here we distinguish. First, is the design phase. ``Design optimization" seeks to answer the question ``what design should we build?" At this stage, there is freedom over parameters including major radius and aspect ratio. For each potential design, a second stage, ``operational optimization" answers the question ``how is this machine best operated?" During the tokamak design process, operational optimization is an internal optimization loop that must be completed to compare the performance of different designs when ideally operated. However, once the tokamak design is fixed, only the latter question remains. During the operational optimization phase, decisions include the plasma impurity mix, the shaping coil currents that determine the plasma shape, and the heating, fueling, and current driven into the plasma. Here, we are specifically interested in the core, steady-state optimization of the second phase, including only parameters that can be changed after the tokamak design has been fixed. The range of operational options available, \textcolor{black}{particularly with regards to the plasma shape, is limited} by decisions made in the design phase. For example, the shape of the vacuum vessel and locations of the poloidal field coils determine what plasma shapes are feasible.

 Here, we consider the plasma impurity composition (effective ion charge: $Z_{eff}$, and main ion fraction: $f_{main}$), plasma pedestal density ($n_{e,ped}$), and plasma geometry (elongation: $\epsilon$, triangularity: $\delta$, and squareness: $\zeta$) as key operational optimization parameters.  A burning plasma, with fusion gain $>$ 5, behaves differently than present experimental plasmas in that it is predominately heated by alpha power, not auxiliary heating. In projected ARC plasmas fusion gain is expected to be much higher then this burning threshold. Therefore, auxiliary heating plays a less crucial role in steady state simulations and will not be considered as an optimization knob. We also elect not to include plasma current as an optimization knob in this work, as currents are tightly coupled with pulse duration, disruption forces, and other engineering constraints. For simplicity, we assume the operator has a choice of the plasma pedestal density.  However, we note that in H-mode it is not necessarily feasible to fuel once a pedestal has been achieved \cite{groebnerElementsHmodePedestal2023, mordijckOverviewDensityPedestal2020}. No additional particles sources are considered. The main ways these optimization parameters impact fusion power performance are through changing the pedestal pressure and heat transport due to turbulence.

\subsection{Impact of Impurity Composition}

There are many different sources of plasma impurities both intentional and intrinsic. Intentionally, impurities are sometimes puffed into the plasma in substantial quantities or laser blow-off systems are used to inject small quantities for diagnostic purposes. Minority species can be added to the plasma, in the case of ARC, hydrogen, for the ICRH scheme \cite{ongenaRecentAdvancesPhysics2017}. Other, inherent sources of impurities include sputtering from the wall \cite{matthewsPlasmaOperationAll2013, wolfrumImpactWallMaterials2017}, sputtering from ICRH antennae \cite{diabMitigationICRFEdge2025}, and helium (alpha particle) ash accumulation \cite{putterichDeterminationTolerableImpurity2019}. 

With so many difference sources, there are a wide variety of impurities in the plasma.  Fortunately, the impact of impurities on turbulent transport mostly be captured by one effective ion charge ($Z_{eff}$) and main ion fraction ($f_{main}$), sometimes referred to as dilution \cite{dominguezImpurityEffectsDrift1993, rodriguez-fernandezEnhancingPredictiveCapabilities2023}. $Z_{eff}$ can be calculated as $Z_{eff} = \sum_i \frac{n_i Z_i^2}{n_e}$, where the sum is taken over all ion species within the plasma. $n_i$ and $Z_i$ are the density and charge of species $i$, and $n_e$ is the electron density. $f_{main}$ can be calculated as $f_{main} = \frac{n_{main}}{n_e}$, where $n_{main}$ is the density of the main ion(s), in the case of ARC, the fuel species deuterium and tritium. Sometimes known as the ``lumped impurity approximation," one impurity species with a charge equivalent to the overall $Z_{eff}$ in concentration that gives the same $f_{main}$ as the overall plasma is used. The additional approximation that the impurity's mass is twice its charge is also commonly used \cite{migliuoloIonTemperatureGradient1992}. 

There are four key ways impurity concentration impacts the amount of steady-state fusion power produced by a reactor: (1) dilution of the fuel species, (2) core plasma turbulent transport, (3) radiative losses, and (4) pedestal height and width. Also of importance, is that as $Z_{eff}$ increases collisionality increases as well decreasing the feasible pulse length for a given design. The net impact of adding impurities can be beneficial; it has previously been noted that $Z_{eff}$ optimization on SPARC L-modes is predicted to increase fusion gain \cite{rodriguez-fernandezCorePerformancePredictions2024}.  The detrimental component of the effect of the dilution of fuel species is easily captured even by 0D modeling: a lower density of fuel ions means a lower fusion power density.  Core plasma transport can either be increased or decreased by the presence of impurities. The main ion ITG (ion temperature gradient) mode, which is predicted to be the primary driver of transport in ARC \cite{howardPerformanceTransportARC2026}, is stabilized by the presence of impurities \cite{weiGyrokineticParticleSimulation2018}. However, impurities are destabilizing to trapped electron modes (TEM) \cite{dominguezImpurityEffectsDrift1993, pusztaiTurbulentTransportImpurities2013}. Additionally, the presence of impurities opens up the possibility of two additional modes: the impurity ITG and the impurity mode \cite{froejdhImpurityEffectsToroidal1993, hanQuasilinearTransportAnalysis2025, migliuoloIonTemperatureGradient1992}. 

Radiative losses, due to line, Bremsstrahlung, and synchrotron radiation, increase with increasing $Z_{eff}$ \cite{freidbergPlasmaPhysicsFusion2007}. There are both negative and positive impacts of increased radiative losses, which can be seen through a simple power balance analysis. In steady state, the energy put into the plasma must be equivalent to the energy coming out of it. Therefore, the energy deposited in the plasma by auxiliary heating, ohmic heating, and alpha heating must be equivalent to the power lost to the divertor through conduction and the power lost through radiation. In a worst case scenario, if the radiative losses become greater than the total power input into the plasma, the plasma cannot be in steady state and would undergo radiative collapse. In a less severe scenario, the radiated power can reduce the amount of power available to be conducted to the divertor such that the scrape-off-layer power is less than that of the L-H transition threshold \cite{martinPowerRequirementAccessing2008}. However, often too much scrape-off-layer power is the concern because it can damage the divertor, which is where impurities can be enormously beneficial. More impurities radiating in the core can mitigate the heat flux seen by the divertor. Additionally, intentional puffing of low charge impurities that radiate in the edge, some of which end up in the core \cite{kallenbachDivertorEnrichmentRecycling2024}, can help achieve good performance in metal-walled devices \cite{wolfrumImpactWallMaterials2017} and to achieve detachment \cite{bodySimpleAccurateModel2025}.

There is substantial experimental evidence that confinement can benefit from the presence of impurities. Higher performance than normally would be expected in L-mode due to increased impurities has been found on TEXTOR-94 \cite{tokarEvidenceSuppressionITGinstability1999}, TEXTOR \cite{unterbergRadiativeImprovedMode2005}, C-Mod \cite{greenwaldModeConfinementAlcator1997, enneverEffectsDilutionTurbulence2015}, TFTR \cite{hillTestsLocalTransport1999},  DIII-D \cite{mckeeImpurityInducedSuppressionCore2000, jacksonEffectsImpuritySeeding2002}, and ASDEX-Upgrade \cite{fableHighconfinementRadiativeLmodes2021}. Sometimes this enhanced L-mode is referred to as ``radiatively improved" mode, a review of which is found in \cite{tokarConfinementMechanismsRadiatively1999}. Recent experimental work in at DIII-D in the negative triangularity regime has also demonstrated confinement improvement with increased impurities \cite{casaliAchievementHighlyRadiating2025}. Improvements in confinement in H-mode, due to a mixture of both core turbulence stabilization and enhanced pedestal performance, are found in DIII-D \cite{osborneEffectPlasmaShape2000, xuEnhancedHmodeBoron2025}, ASDEX Upgrade \cite{dunneGlobalPerformanceEnhancements2016}, JT-60U \cite{uranoRolesArgonSeeding2015}, and JET \cite{huberPeculiarityHighlyRadiating2020}. 

\subsection{Impact of Geometric Shaping}

\begin{figure}
    \centering
    \includegraphics[width=0.5\linewidth]{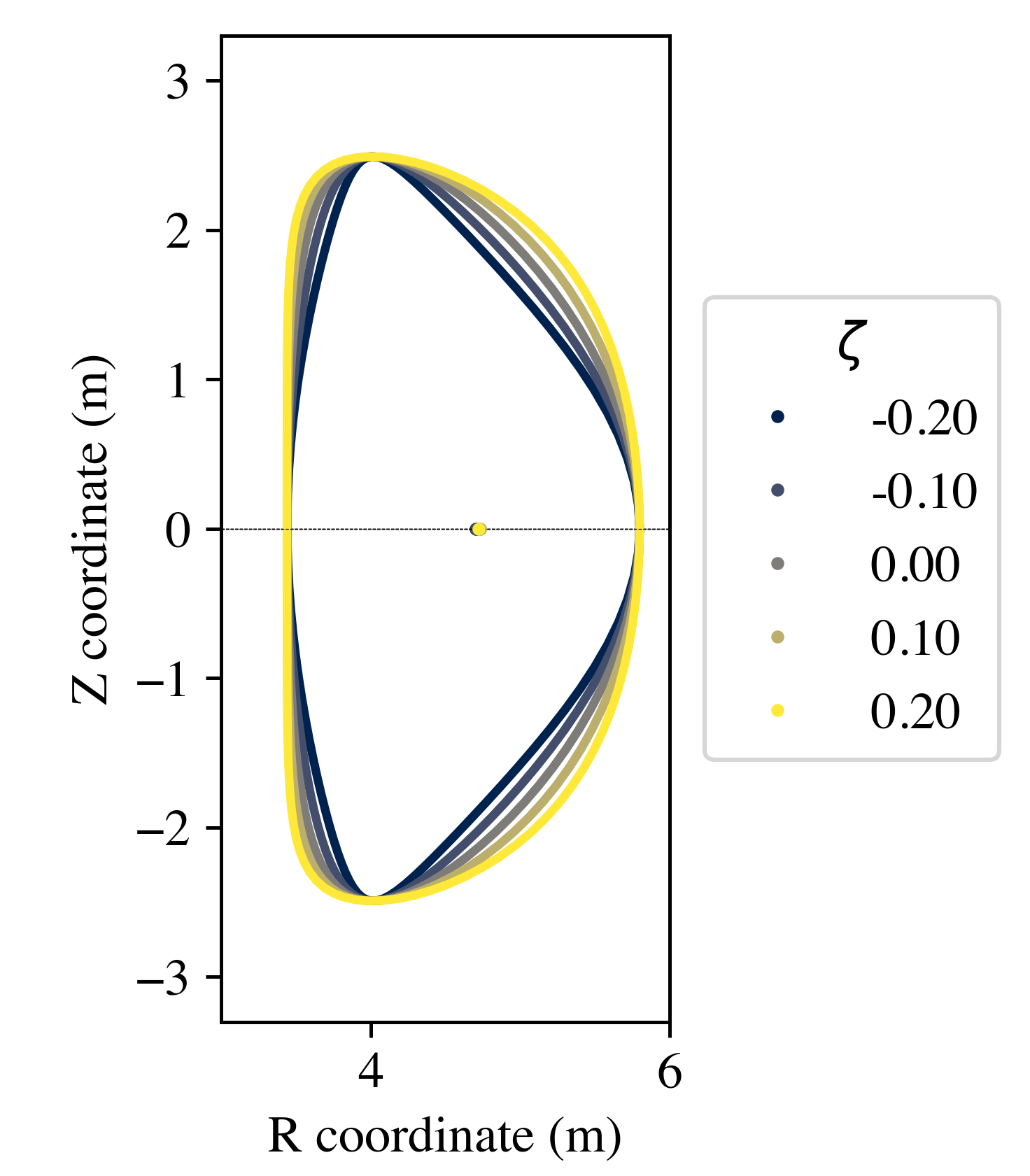}
    \caption{Scan of squareness for equilibria with the same aspect ratio, major radius, $\kappa$, and $\delta$ as ARC V3A.}
    \label{fig:squareness_scan_equilibria}
\end{figure}

The benefits of using non-circular poloidal cross sections, with shaping, has long been appreciated on tokamaks worldwide \cite{lomasVariationConfinementElongation2000, osborneEffectPlasmaShape2000, laoDependenceEdgeStability2001}. While elongation and triangularity are regularly adjusted in experiments, control of the higher order shaping parameter of squareness has be experimentally demonstrated \cite{holcombOptimizingStabilityTransport2009, kolemenPlasmaModellingResults2011, liuEffectsPlasmaSquareness2026, chenPlasmaShapeOptimization2019, degraveMagneticControlTokamak2022}. Equilibria resulting from a scan of squareness are shown in Figure \ref{fig:squareness_scan_equilibria}. A widely used model to characterize tokamaks' plasma shape is the Miller parameterization \cite{millerNoncircularFiniteAspect1998}, which includes aspect ratio, elongation, and triangularity. It was then expanded to the Turnbull-Miller parameterization, which adds the variable of squareness \cite{turnbullImprovedMagnetohydrodynamicStability1999}, the geometric definition of which can be found in \cite{turnbullImprovedMagnetohydrodynamicStability1999, holcombOptimizingStabilityTransport2009}. Recently, the Miller Extended Harmonic (MXH) parameterization was put forth \cite{arbonRapidlyconvergentFluxsurfaceShape2020}. While the two earlier parameterizations relied upon geometric definitions of the input parameters, the MXH model uses a Fourier harmonic approach\footnote{Note the s2 coefficient referred to as squareness in the MXH parametrization is not exactly equivalent to the geometric definition.}. In this work, the MXH parameterization is used. 

Plasma shaping has the strongest effects in the outermost flux surfaces, which then get rounded out moving to the core. Therefore, there are strong impacts on the pedestal from shaping. Ballooning modes are pressure gradient driven MHD instabilities that occur in the unfavorable magnetic curvature region present in toroidal devices. Peeling modes are current-driven MHD instabilities naturally driven in the pedestal due to the high bootstrap currents induced by the pedestal gradient. Elongation is known to primarily stabilize the ballooning modes, and triangularity is known to stabilize the peeling modes, while slightly destabilizing the ballooning modes \cite{kimDifferentRolesElongation2022}.  The effect of squareness on the pedestal pressure are not as clearly understood; although lower (numeric not absolute) values are usually preferred \cite{holcombOptimizingStabilityTransport2009, leonardPedestalPerformanceDependence2007, aibaEffectsSharpnessPlasma2007, parisiPredictionELMfreeOperation2025}, too low of squareness can generate magnetic cusps \cite{turnbullImprovedMagnetohydrodynamicStability1999}.

Increased shaping can also suppress core turbulent transport. Both positive triangularity \cite{pueschelReducingTransportExtreme2024} and elongation \cite{belliEffectsPlasmaShaping2008, angelinoRolePlasmaElongation2009} have been shown to enhance zonal flows. The impact of squareness is less clear \cite{holcombOptimizingStabilityTransport2009, joinerIonTemperatureGradient2010}. Beyond confinement implications, shaping, particularly elongation, also impacts the vertical stability of the plasma \cite{leeTokamakElongationHow2015, menardFusionNuclearScience2016}, so too high of elongation values are not feasible. \textcolor{black}{Achieving higher elongation than the nominal ARC V3A design requires additional vertical stabilization capabilities \cite{leutholdARCPhysicsBasis2026}, which could have additional monetary cost.} Moreover, adjustments to the geometry can allow the location of the burning plasma region to be moved to higher volume regions in plasmas, for example, squareness when a high Shafranov shift is present \cite{parisiDoublingFusionPower2025}.

While thus far the implications of impurity composition and geometric shaping have been discussed separately, they have joint effects. In transitioning from carbon to metal walls, different pedestal effects were found between low and high triangularity \cite{wolfrumImpactWallMaterials2017, beurskensEffectMetalWall2013}. The $Z_{eff}$ itself was found to have a dependence on triangularity \cite{maggiLHPowerThreshold2014}. On ASDEX Upgrade, impurity seeding was found to primarily impact the pedestal temperature while triangularity primarily impacted the pedestal density \cite{dunneGlobalPerformanceEnhancements2016}. The pedestal density determines whether a plasma pedestal is limited by a peeling-mode or a ballooning-mode; the density at which this transition occurs depends both on plasma composition and geometry. Therefore, it is critical to optimize impurity composition, plasma geometry, and pedestal density simultaneously. 

In the following section, the methods used in this work will be discussed. Section \ref{sec:scans} will provide intuition regarding the impact of individual and pairs of operational optimization parameters on the plasma performance. In section \ref{sec:BO}, we apply Bayesian optimization techniques to find optima allowing for variation in all parameters considered in this work.  Section \ref{sec:discussion} discusses conclusions and future work.

\section{Modeling Methodology}
\label{sec:methods}

\subsection{\texttt{MAESTRO} setup}

Modeling in this work was performed with \texttt{MAESTRO} \cite{rodriguez-fernandezAcceleratingIntegratedModeling2026}, a capability of the open source software, \texttt{MITIM-fusion} \cite{rodriguez-fernandezPabloprfMITIMfusion2026}. It takes as inputs engineering parameters and makes plasma performance predictions with self-consistent heating profiles, radiation losses, H-mode boundary conditions, and core transport. The equilibrium is calculated in this work by scaling up and down the Miller parameterization of the nominal ARC V3A design. The initial guess of the plasma normalized pressure ($\beta_N$) is 2.0. The equilibrium is then passed to \texttt{TRANSP} \cite{pankinTRANSPIntegratedModeling2025} to allow the current distribution to relax for 20 seconds of plasma modeling time. After this, self consistent heating profiles and equilibria, including the effects of fast ion pressure, are calculated via \texttt{TRANSP} using \texttt{NUBEAM} \cite{pankinTokamakMonteCarlo2004}, \texttt{TORIC} \cite{brambillaNumericalSimulationIon1999}, and \texttt{FPPMOD} \cite{hammettFastIonStudies1986}. 

The boundary condition at the top of the H-mode pedestal is calculated with \texttt{EPED} \cite{snyderDevelopmentValidationPredictive2009, snyderFirstprinciplesPredictiveModel2011}. The equilibrium is passed to \texttt{EPED} here using the MXH parameterization \cite{MXH_EPED_personal_communication}. Nominally the assumed pedestal density input into \texttt{EPED} is $2.1 \times 10^{20}/m^3$, with a ratio between the separatrix density and the pedestal density ($n_{sep}/n_{ped}$) of 0.4. It is assumed that ion and electron temperature at the top of the pedestal are equal. 

The location of the boundary condition is set in accordance with the predicted pedestal width. Transport inside this boundary is calculated using the steady-state, \texttt{PORTALS} transport solver \cite{rodriguez-fernandezEnhancingPredictiveCapabilities2024, rodriguez-fernandezNonlinearGyrokineticPredictions2022}. Turbulent transport is predicted using \texttt{TGLF} \cite{staeblerTheorybasedTransportModel2007} \texttt{SAT2} \cite{staeblerVerificationQuasilinearModel2021} with a maximum of 6 basis functions (\texttt{nbasis\_max = 6}). Neoclassical transport is calculated via \texttt{NEO} \cite{belliKineticCalculationNeoclassical2008}.  The electron particle flux, electron heat flux, and ion heat flux channels are predicted; impurity transport is not predicted in the scope of this study. The fuel species is 50\% deuterium and 50\% tritium, which are modeled as separate species in this use of \texttt{TGLF}. The impurities are tungsten, at a concentration of $1.5 \times 10^{-5} \times n_e$, hydrogen at a concentration of $0.05 \times n_e$, and a lumped low-Z impurity species with a charge and concentration consistent with the $Z_{eff}$ and $f_{main}$ being modeled. Helium ash could be thought of as being one impurity, among others, included in the low-Z impurity, but it is not accounted for self-consistently with the fusion reaction rate. In the nominal case, $Z_{eff} = 1.5$ and $f_{main} = 0.85$, which results in a lumped impurity of charge $Z_{imp} = 4.4$ and concentration of $f_{imp} =$ 3\%. All concentrations are with respect to the electron density ($n_e$). Radiation is modeled with all four ion species and the electrons using a Chebyshev fit to the ADAS database \cite{rodriguez-fernandezNonlinearGyrokineticPredictions2022, candyTokamakProfilePrediction2009}. The lumped impurity is treated as the element with the closest atomic charge to the calculated lumped charge state. The radial concentration profiles are calculated assuming a approximately flat $Z_{eff}$ profile of the stated value in the core, with higher $Z_{eff}$ at the outermost radii.  Iteration between \texttt{TRANSP}, \texttt{EPED}, and \texttt{PORTALS-TGLF} continue until the simulation is converged. \textcolor{black}{Pulse duration calculations are made using the functionality within \texttt{CFSPOPCON} \cite{bodyCfsenergyCfspopconV7022024}, which leverages the model found in \cite{barrPowerbalanceModelLocal2018}. Since no ramp up optimization is performed, the stated values should be treated as relative. However, this model does capture the dependence of the plasma resistance to the $Z_{eff}$, temperature, and elongation.} The fusion power predicted for the nominal ARC V3A design with these exact modeling assumptions is 1140 MW.  This value is at the upper end of the predicted fusion power in \cite{howardPerformanceTransportARC2026} due to the parameterization being used to approximate the equilibrium to facilitate scans over different geometries. While \texttt{MAESTRO} is a relatively new framework, the models within it are all extensively validated, and comparisons of predictions made with full \texttt{MAESTRO} framework against three experimental papers were done in \cite{yannaMultiTokamakAssessmentModeled2025}.

\section{Operating space scoping with 1D and 2D scans} 
\label{sec:scans}

First, we perform brute force scans of 2D cross sections of the operating space and 1D stand alone scans with \texttt{TGLF} and \texttt{EPED} to understand the location of the maxima and overall structure of the optimization space. 

\subsection{Composition}
\begin{figure}[h]
    \centering
    \begin{subfigure}{0.49\linewidth}
        \includegraphics[width=\linewidth]{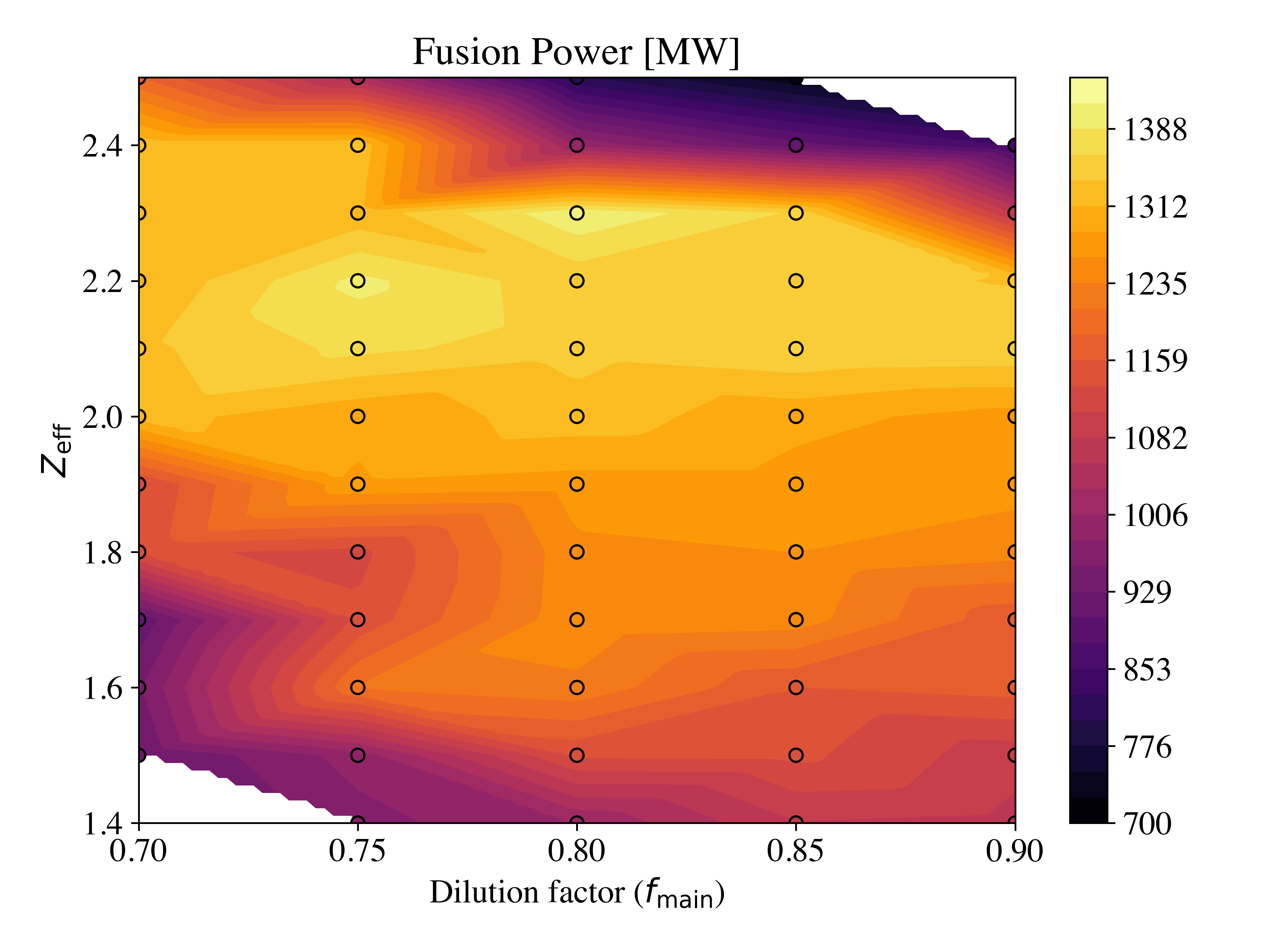}
        \caption{}
        \label{subfig:comp_scan_fusion_power}
    \end{subfigure}
    \begin{subfigure}{0.49\linewidth}
        \includegraphics[width=\linewidth]{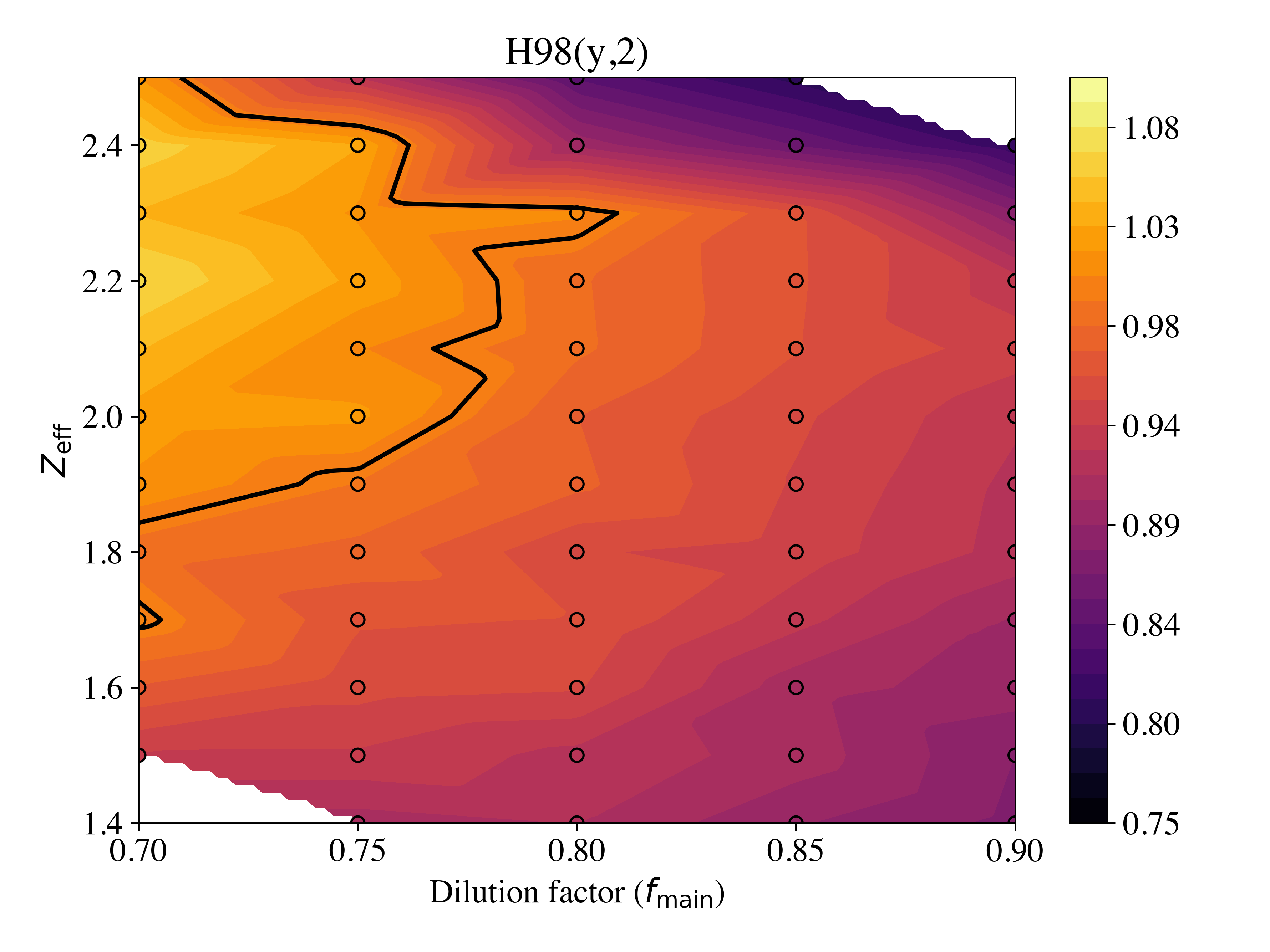}
        \caption{}
        \label{subfig:comp_scan_H98}
    \end{subfigure}
    \caption{The resulting (a) fusion power and (b) scalar multiplicative factor on the energy confinement time predicted by the ITER98(y,2) \cite{iterphysicsexpertgrouponconfinementandtransportChapter2Plasma1999} from performing a two dimensional scan of $Z_{eff}$ and $f_{main}$. The circles show the locations of MAESTRO simulations performed with interpolation between them. The nominal operating point for ARC V3A has $Z_{eff} = 1.5$ and $f_{main}=0.85$, depicted in the lower right side of these plots. By going to an operating point with a higher $Z_{eff}$  $\sim$25\% more fusion power can be achieved. The variation in $H98_{(y,2)}$ depicted in (b) demonstrates that capturing the full effects of changing impurity composition requires physics-based models. The black line denotes the $H98(y,2)$ = 1 surface, inside of which, 0D models would give relatively pessimistic predictions.}
    \label{fig:comp_scan}
\end{figure}

\begin{figure}[h!]
    \centering
    \includegraphics[width=\linewidth]{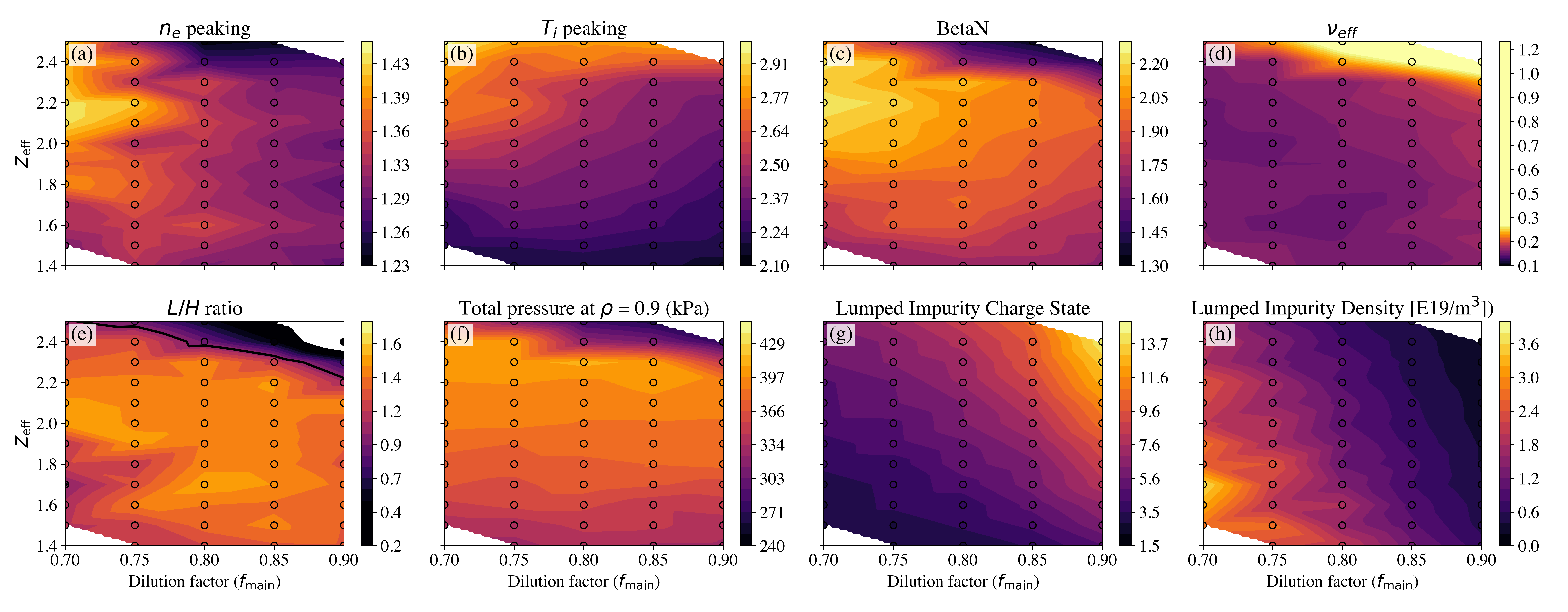}   
    \caption{Additional quantities of interest for the impurity composition scan, varying $Z_{eff}$ and $f_{main}$, also depicted in Figure \ref{fig:comp_scan}; \texttt{MAESTRO} simulations performed are circled in black. Subplots (a) and (b) show the electron density and ion temperature peaking respectively, where the peaking is the ratio between the on axis and volume averaged values. Subplot (c) shows the value of volume averaged $\beta_N$. Subplot (e) shows the effective collisionality as defined \cite{angioniScalingDensityPeaking2007} and multiplied by $Z_{eff}/2$ to relax the assumption of $Z_{eff} = 2.0$ made in that paper. Subplot (e) shows the ratio between the power threshold predicted by the scaling in \cite{martinPowerRequirementAccessing2008} and the scrape-off-layer power. Subplot (f) shows the plasma pressure of $\rho = 0.9$, which is a proxy for the pressure at the top of the pedestal as given by EPED (although width effects are neglected). Subplots (g) and (h) show the conversion from $Z_{eff} - f_{main}$ space to that of the lumped impurity charge state and density.}
    \label{fig:comp_scan2}
\end{figure}

We conduct 2D scans in plasma composition and geometry separately to understand the underlying physics before doing the full operational optimization. We begin by scanning impurity composition, varying $Z_{eff}$ and $f_{main}$, at fixed $n_{e,ped}$ and plasma geometry. The resulting fusion power from this scan can be found in Figure \ref{subfig:comp_scan_fusion_power} with additional quantities plotted in Figure \ref{fig:comp_scan2}. The nominal impurity composition for ARC V3A has $Z_{eff} = 1.5$ and $f_{main} = 0.85$. By optimizing only impurity composition the predicted fusion power increases by $\sim$25\%. The mapping between $Z_{eff}$ and $f_{main}$ and lumped impurity charge state and density scan be found in subplots (g) and (h) of Figure \ref{fig:comp_scan2}. The optimal fusion power is achieved when the lumped impurity has a higher charge (and mass) as well as there being a higher density of the lumped impurity. At too high of $Z_{eff}$ the fusion power decreases sharply. It is also observed that there is some interdependence between $Z_{eff}$ and $f_{main}$ in the fusion power they produce, with a stronger, but still relatively weak, dependence of $f_{main}$ at lower values of $Z_{eff}$.

Before we look deeper into the reasons for this response of fusion power to impurity composition, there is another important result here. One can think of $H98_{(y,2)}$ as a multiplicative ``fudge-factor" that would have to be applied to the results of the ITER98(y,2) scaling law \cite{iterphysicsexpertgrouponconfinementandtransportChapter2Plasma1999} to recover the same confinement as is predicted by the MAESTRO simulations. In 0D modeling, using the energy confinement time scaling with a constant value of $H98_{(y,2)}$ causes the fusion power to strictly decrease with the main ion fraction as there is a lower fuel density. However, while for the nominal values of $Z_{eff}$ and $f_{main}$ the ITER98(y,2) scaling law over-predicts the energy confinement time by 10\%, it actually under-predicts the energy confinement time for cases with higher $Z_{eff}$ and lower $f_{main}$. The ITER98(y,2) scaling has a root mean square error of 14.5\%. The variation in $H98_{(y,2)}$ seen due to changing impurity composition therefore spans about one standard deviation of the scaling law. Capturing the impact of impurity composition requires physics-based models; 0D models can give pessimistic predictions for operating points with high $Z_{eff}$ and low $f_{main}$. 

There are four key ways that the impurity composition changes the fusion power. First,  a lower $f_{main}$ reduces the density of the fuel ions, an effect easily captured by both 0D and physics-based modeling. In Appendix A, Figures \ref{fig:profiles_scan_Zeff_1.5}d and \ref{fig:profile_scan_Zeff_2.2}d depict the reduced density of fuel ions at lower $f_{main}$.  Since fusion power scales as the fuel density squared, this is a large effect dependent only on $f_{main}$. 

\begin{figure}
    \centering
    \begin{subfigure}{0.8\linewidth}
        \includegraphics[width=\linewidth]{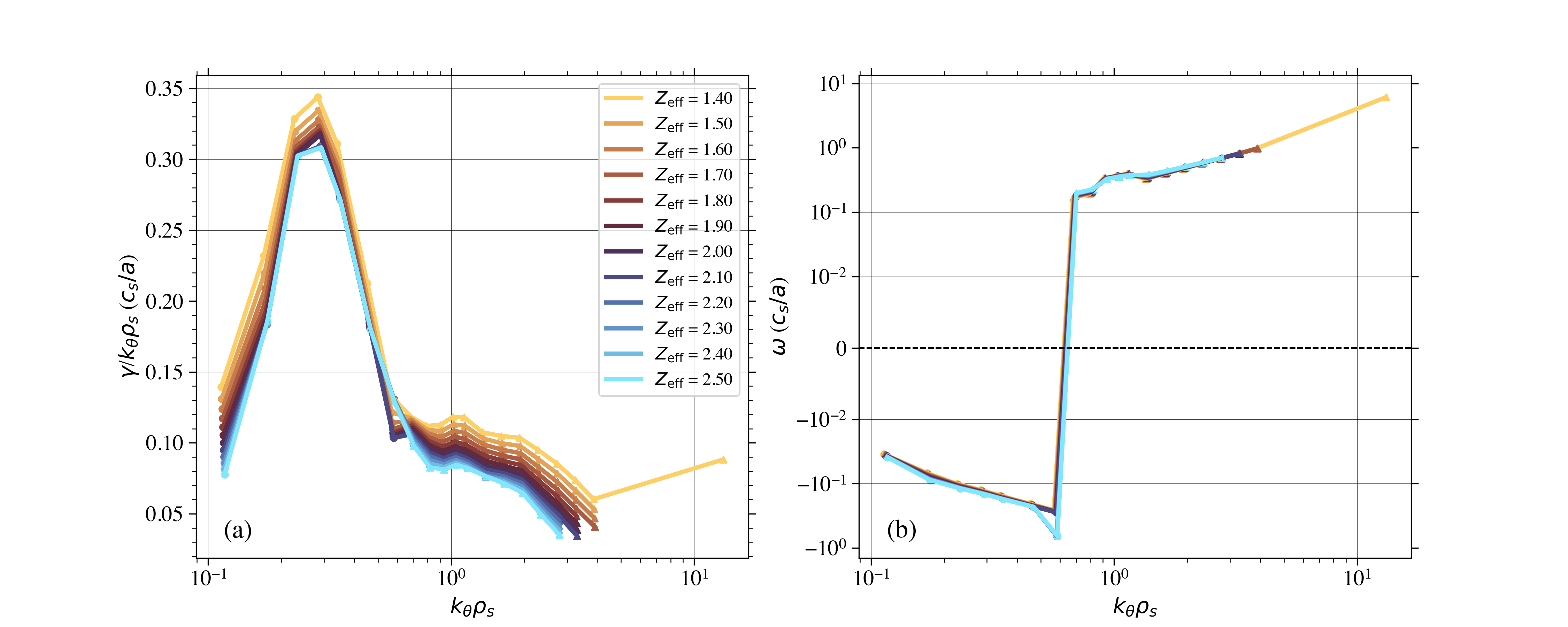}
    \end{subfigure}
    \hfill
    \begin{subfigure}{0.8\linewidth}
        \includegraphics[width=\linewidth]{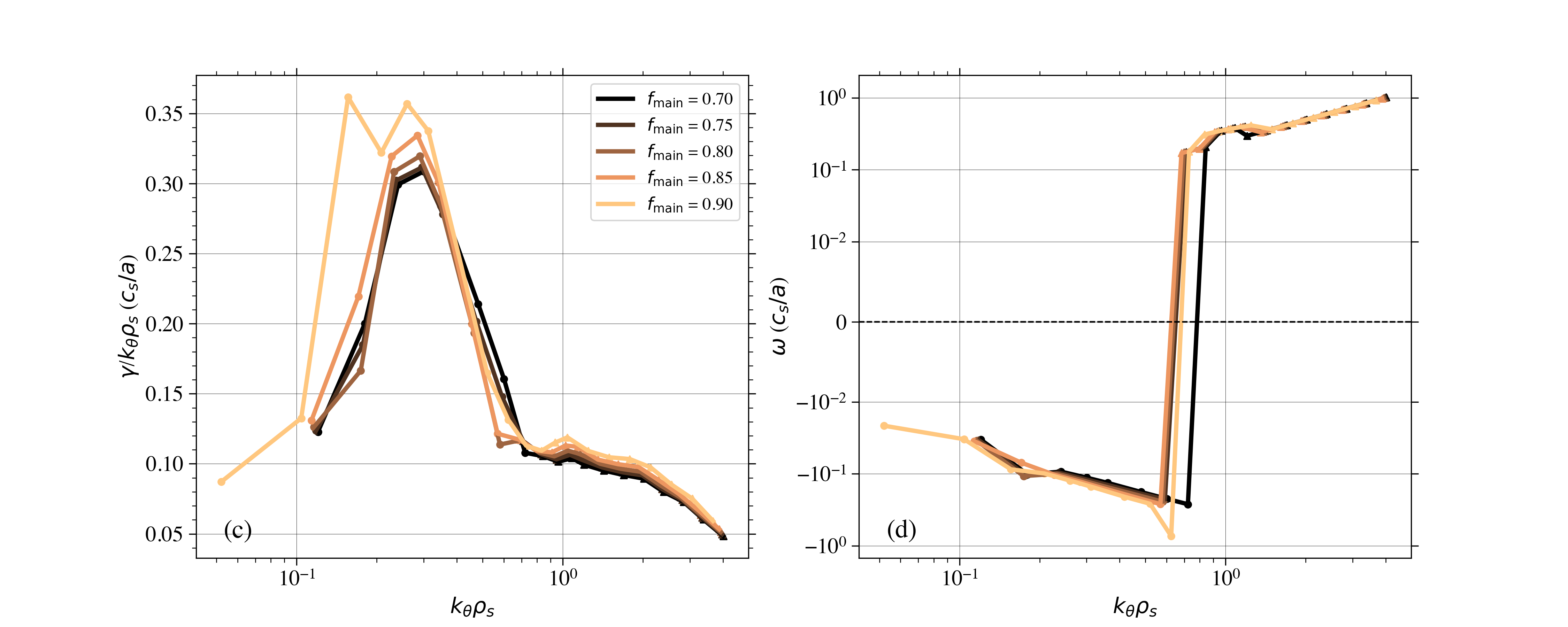}
    \hfill
    \end{subfigure}
        \begin{subfigure}{0.8\linewidth}
        \includegraphics[width=\linewidth]{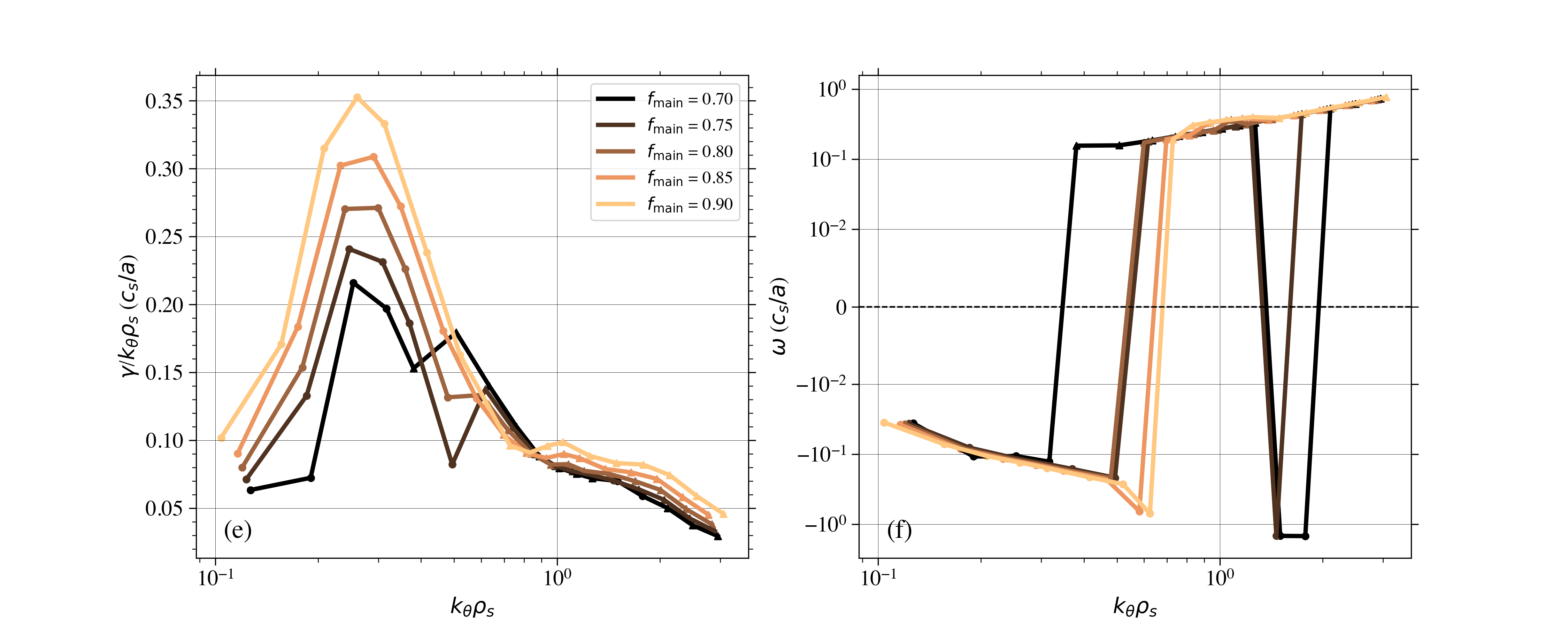}
    \end{subfigure}
    \caption{Stand-alone \texttt{TGLF} scans at $\rho = 0.6$ to assess the linear impact on turbulence of changing $Z_{eff}$ at nominal $f_{main}$ (a, b), $f_{main}$ at nominal $Z_{eff}$ = 1.5 (c,d), and $f_{main}$ at $Z_{eff} = 2.2$. The normalized growth rates are shown in (a) and (c), and the real frequencies, with a negative ion direction, are shown in (b) and (d).  Both increasing $Z_{eff}$ and decreasing $f_{main}$ have a stabilizing effect. At higher values of $Z_{eff}$, lowering $f_{main}$ has a much stronger stabilization effect than at the nominal value of $Z_{eff}$.}
    \label{fig:TGLF_composition_stand_alone}
\end{figure}

Second, the presence of impurity ions, increasing $Z_{eff}$ and decreasing $f_{main}$, can stabilize the main ion ITG, which is the primary driver of transport. This is directly observed in standalone \texttt{TGLF} scans, seen in Figure \ref{fig:TGLF_composition_stand_alone}. They are initialized with the output of the nominal ARC V3A \texttt{MAESTRO} simulation used in this work. The lumped impurity charge (and corresponding atomic mass) as well as the lumped impurity concentration were varied to scan $Z_{eff}$ and $f_{main}$ separately. These adjustments are made outside of \texttt{MAESTRO}; there are no changes in background values, background gradients, source terms, or pedestal effects included in these scans. Changing $Z_{eff}$ at the nominal $f_{main}$, shown in Figures \ref{fig:TGLF_composition_stand_alone}a and \ref{fig:TGLF_composition_stand_alone}b, results in stabilization across the turbulence spectrum including in both the ITG and ETG portions. Next, we try changing $f_{main}$ while holding $Z_{eff}$ constant. Since on SPARC, the effect of changing $f_{main}$ was found to have a dependence on the value of $Z_{eff}$ in \cite{rodriguez-fernandezStudiesImpurityinducedTurbulence2025}, the nominal and optimal values of $Z_{eff}$ were tested here. ITG stabilization is observed when $f_{main}$ decreases for both $Z_{eff} = 1.5$ and $Z_{eff} = 2.2$. However the effect is much stronger at higher $Z_{eff}$, which can be seen in Figures \ref{fig:TGLF_composition_stand_alone}c and \ref{fig:TGLF_composition_stand_alone}e. 

Third, changing impurities results in changes to the source/sink terms in the steady state power balance. Beyond feedback loops of other beneficial effects increasing fusion power, an increased presence of impurities increases the amount of radiated power. More radiated power means, for the same input power, less heat flux to the scrape-off-layer decreasing the temperature gradients in the stiff, core plasma. If the scrape-off-layer power gets too low, the L-H transition threshold power may not be met. If the scrape-off-layer power goes negative, there is no steady-state solution and radiative collapse would have occurred. However, this also means, beneficially, the divertor sees less heat flux.  

\begin{figure}[h]
    \centering
    \includegraphics[width=0.7\linewidth]{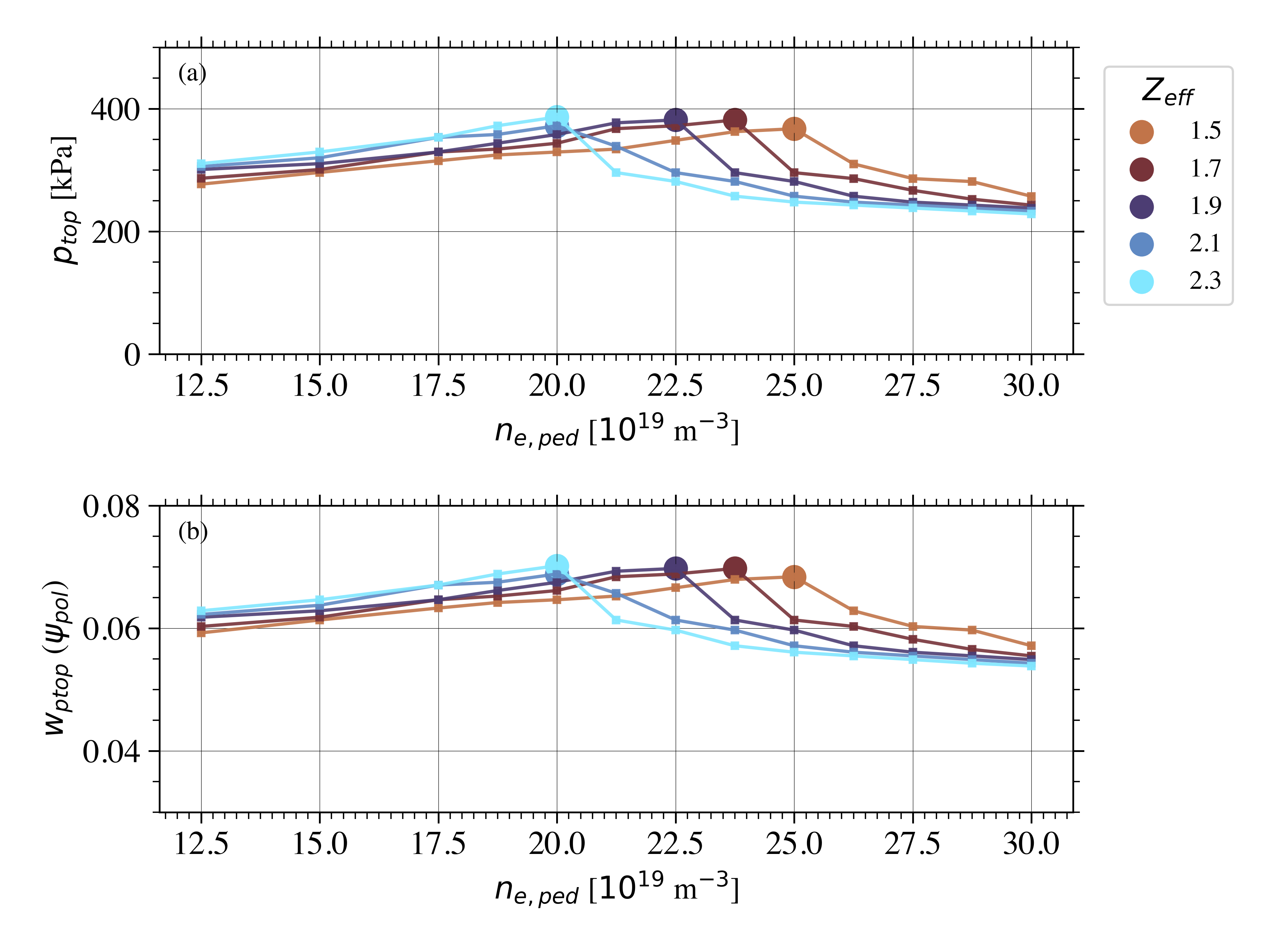}
    \caption{Impact of $Z_{eff}$ on the pedestal height ($p_{top}$) and width ($w_{ptop}$) at different pedestal densities ($n_{e,ped}$). Higher $Z_{eff}$ causes ballooning modes to begin limiting the pedestal at lower densities. However, for the same density, at higher $Z_{eff}$, the peeling branch pressure is slightly elevated.}
    \label{fig:EPED_Zeff_scan}
\end{figure}

\begin{figure}
    \centering
    \includegraphics[width=0.7\linewidth]{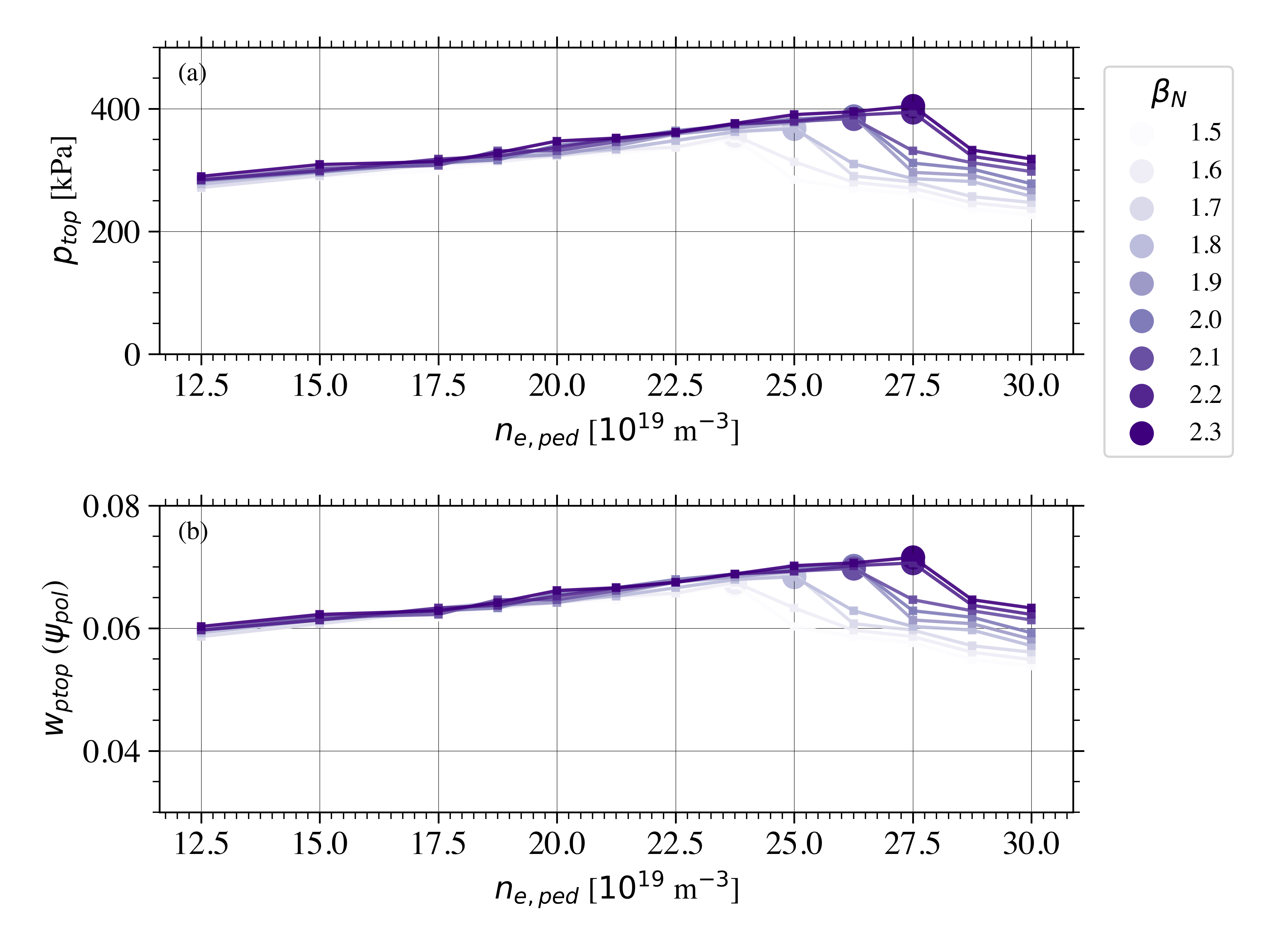}
    \caption{Impact of $\beta_N$ on the pedestal height ($p_{top}$) and width ($w_{ptop}$) at different pedestal densities ($n_{e,ped}$). Higher $\beta_N$ allows slight increases the pressure on the peeling branch, for the same value of $n_{e,ped}$ as well as increasing the density at which ballooning modes begin to the limit the pedestal.}
    \label{fig:EPED_betaN_scan}
\end{figure}

Fourth, the pressure at the top of the pedestal has a direct dependence on $Z_{eff}$, as well as an indirect dependence on plasma on performance via $\beta_N$. A standalone scan of $Z_{eff}$ performed with \texttt{EPED} is shown in Figure \ref{fig:EPED_Zeff_scan}. All other \texttt{EPED} input parameters are held constant, including the nominal plasma shape (with $\zeta = 0$) and a $\beta_N$ of 1.8. Therefore, only the direct effect of changing $Z_{eff}$ is captured, not the indirect effects of plasma performance via $\beta_N$.  When the pressure at the top of the pedestal is plotted against $n_{e,ped}$ it first increases and then decreases. The lower $n_{e,eped}$ portion of the plots, where pressure increases with $n_{e,ped}$, is known as the peeling branch. The higher $n_{e,ped}$, when pressure decreases with $n_{e,ped}$, is known as the ballooning branch. Increasing $Z_{eff}$ increases the pedestal pressure for a given value of $n_{e,ped}$, consistent with the impact of increased collisionality on the bootstrap current. However, increasing $Z_{eff}$ destabilizes ballooning modes, reducing the density at which ballooning modes begin to limit the pedestal. Therefore, the maximum possible pedestal pressure is relatively constant despite changing $Z_{eff}$. The indirect impact of $f_{main}$ via $\beta_N$ can be observed in Figure \ref{fig:EPED_betaN_scan}. Here, plasma shape and composition are held to their nominal values while $\beta_N$ is scanned. As $\beta_N$ increases, the pressure on the peeling branch increases for a given value of $n_{e,ped}$. More notably, the density at which ballooning modes limit the pedestal also increases. These changes are due to the effect of the Shafranov shift. Therefore, increasing plasma $\beta_N$ increases the pedestal height, thereby further increasing $\beta_N$, which leads to a virtuous cycle. This effect means the pedestal performance depends on the core performance, which is why working with coupled core-pedestal models is critical for this optimization. 

Now we consider how all these effects combine to affect plasma performance. The sudden drop off in fusion power, for constant $n_{e,ped}$ at high $Z_{eff}$, seen in Figure \ref{fig:comp_scan}a is due to the pedestal becoming ballooning limited. This can be verified by looking at Figure \ref{fig:comp_scan2}f, which shows a dramatic decrease in the pedestal pressure at this point. It can also been seen in scans with an \texttt{EPED-NN} \cite{hallMultiFidelityPredictiveCore2024} are found in the appendix in Figure \ref{fig:profile_scan_fmain_0.85}h. There is also a contribution of the cyclical effect of $\beta_N$, which also decreases notably for too high of values of $Z_{eff}$. Whether there is access to H-mode is also an important question when using \texttt{EPED} predictions. The stated error on L-H transition power threshold scaling law found in \cite{martinPowerRequirementAccessing2008} is 30\%.  It is confirmed in Figure \ref{fig:comp_scan2}e that it is reasonable to expect access to H-mode as $f_{LH} > 1.3$ for reasonable values of $Z_{eff}$, despite increased radiation with increasing $Z_{eff}$.  While the H-mode power threshold threshold has experimentally been shown to increase with $Z_{eff}$ \cite{takizukaRolesAspectRatio2004, maggiLHPowerThreshold2014, bourdelleModeTransitionRole2014}, there is still substantial margin. 

Figure \ref{fig:comp_scan2}b shows the ion temperature peaking, which increase in part due to ITG stabilization. It has been shown both experimentally \cite{angioniScalingDensityPeaking2007} and computationally \cite{howardHighFidelityPredictions2025} density peaking has a strong dependence on the collisionality. Indeed, the density peaking shown in Figure \ref{fig:comp_scan2}a increases with decreasing effective collisionality, shown in Figure \ref{fig:comp_scan2}d. These more peaked profiles, for the same auxiliary power, result in the increased value of $H98_(y,2)$, seen in Figure \ref{fig:comp_scan}b, with both increasing $Z_{eff}$ and decreasing $f_{main}$. The relatively constant value of fusion power with changing $f_{main}$, seen in Figure \ref{fig:comp_scan}a, suggests in reducing $f_{main}$ the effects of reduced fuel density and increased ITG stabilization cancel each other out.

\subsection{Plasma Geometry}
\label{subsec:geometric_shaping}

\begin{figure}[h]
    \centering
    \begin{subfigure}{0.49\linewidth}
        \includegraphics[width=\linewidth]{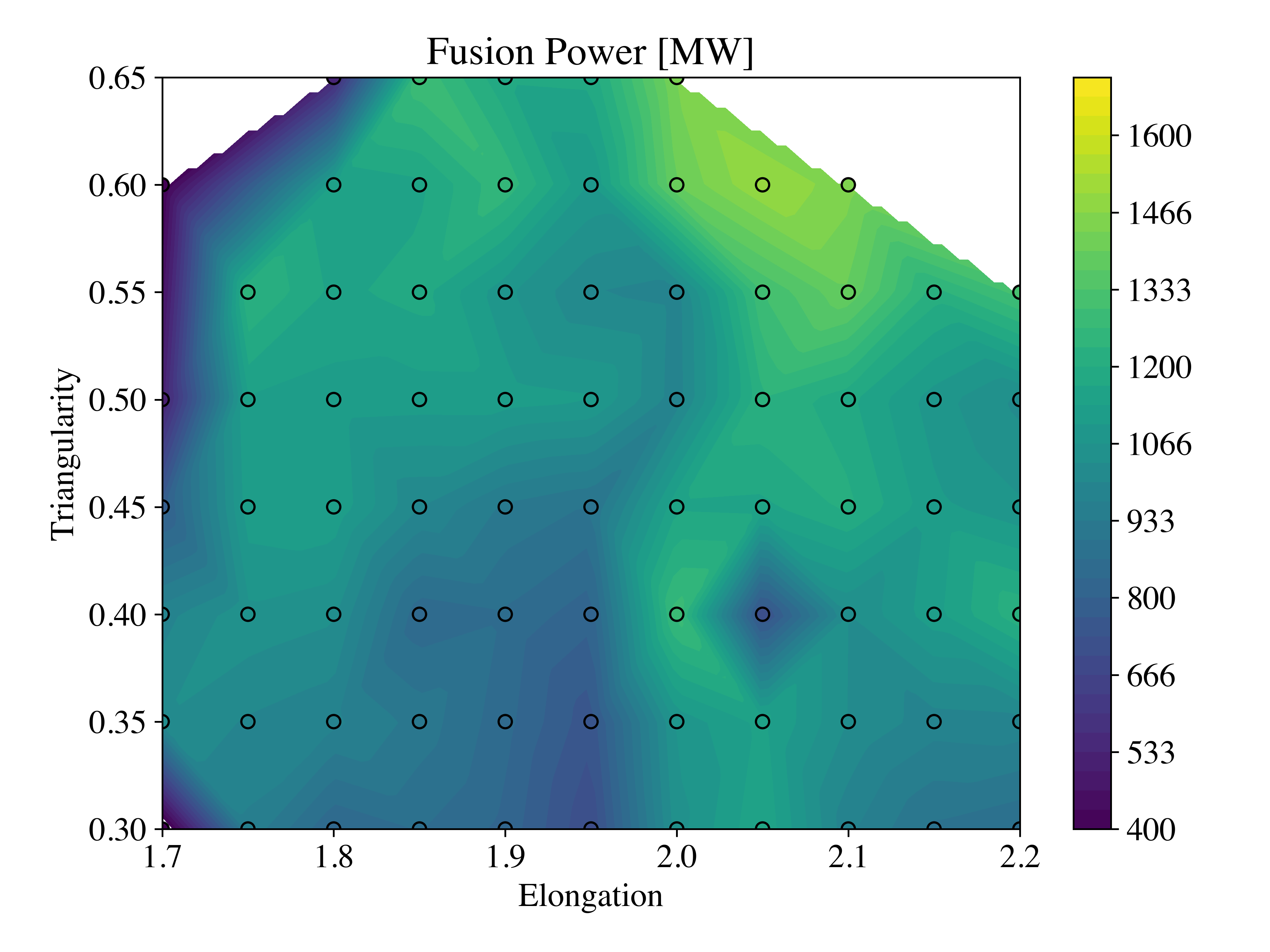}
        \caption{}
        \label{subfig:Pfus_kappa_vs_delta}
    \end{subfigure}
    \begin{subfigure}{0.49\linewidth}
        \includegraphics[width=\linewidth]{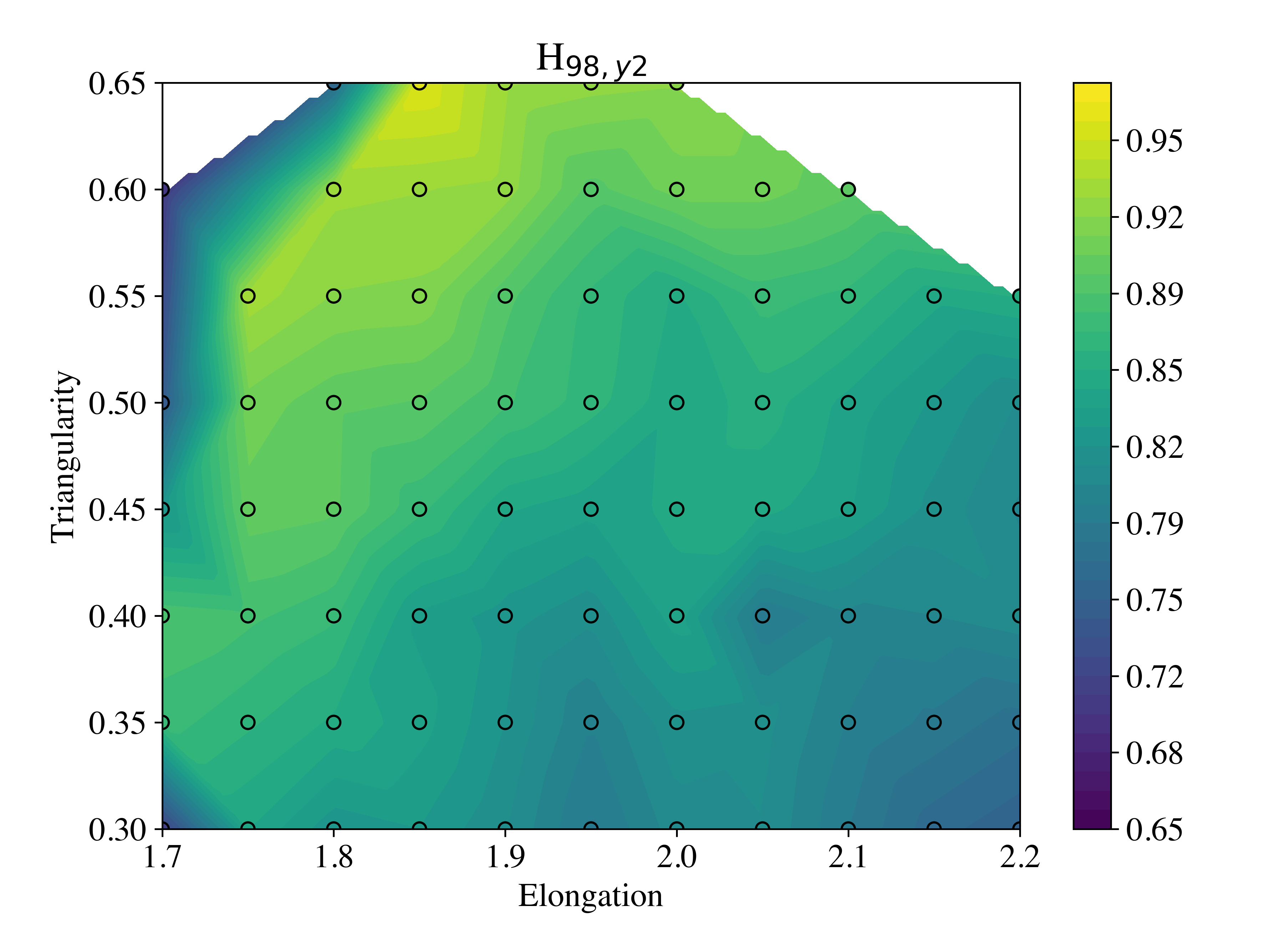}
        \caption{}
        \label{subfig:H98_kappa_vs_delta}
    \end{subfigure}
    \caption{The resulting (a) fusion power and (b) scalar multiplicative factor on the energy confinement time predicted by the ITER98(y,2) scaling \cite{iterphysicsexpertgrouponconfinementandtransportChapter2Plasma1999} from performing a two dimensional scan of elongation ($\kappa$) and triangularity ($\delta$). The nominal values at the 99.5\% flux surface are $\kappa = 1.77$ and $\delta = 0.49$. The circles show the locations of \texttt{MAESTRO} simulations performed with interpolation between them. By going to an operating point with stronger shaping, still enforcing $\kappa < 2.0$, $\sim$20\% more fusion power can be achieved. The variation in $H98_{(y,2)}$ depicted in (b) demonstrates that changes in shaping are not fully captured by the empirical scaling laws. The benefit of increasing elongation is too strong in the ITER98(y,2) scaling law. While triangularity is not included in this scaling law, increased $\delta$ subtly increases the energy confinement time.}
    \label{fig:shaping_scan1}
\end{figure}

Next, we explore the effects of plasma shaping. We begin by scanning $\kappa$ and $\delta$ , using \texttt{MAESTRO} to make performance predictions, while holding $n_{e,ped}$, $Z_{eff}$, $f_{main}$, and $\zeta$ fixed. The nominal values for the ARC V3A design at the 99.5\% flux surface are $\kappa = 1.77$ and $\delta = 0.49$. Throughout this paper, whenever $\kappa$, $\delta$, and $\zeta$ are discussed, they refer to the value at the $\rho = 99.5$\% flux surface. For these scans, the triangularity was limited to $\delta <0.65$ for feasibility with reasonable shaping coils. While $\kappa \lesssim 2.0$ is likely necessary for vertical stability in an ARC-like device, a broader range was scanned here to better understand the behavior of the whole space. The fusion power predicted in this scan can be seen in Figure \ref{fig:shaping_scan1}a. By increasing both $\kappa$ and $\delta$, fusion power can increase by $\sim$20\% while enforcing $\kappa < 2.0$, and it can increase as much as $\sim$30\% for an optimistic range of allowable $\kappa$. Operation with higher elongation and triangularity than the nominal design is favorable, \textcolor{black}{if possible}, although fusion power decreases with increased elongation at extremely high values. 

Again, the importance of using physics-based models can be seen by considering the value of $H98_{(y,2)}$ that would be required to modifying the ITER98(y,2) scaling to get the same energy confinement time predicted by this modeling. While elongation is included in the energy confinement time scaling, $\tau_E \propto \kappa^{0.78}$, we can see it actually has too strong of a dependence with respect to physics-based modeling. As $\kappa$ increases, the $H98_{(y,2)}$ decreases to compensate for the over prediction of confinement with elongation. Meanwhile, $\delta$ is not included in the ITER98(y,2) scaling, but confinement is observed here to increase with $\delta$. That the highest fusion power occurs for a different geometry than the highest value of $H98_{(y,2)}$ emphasizes that fusion power does not directly depend on confinement enhancement.

\begin{figure}

    \begin{subfigure}{\linewidth}
        \center
        \includegraphics[width=0.8\linewidth]{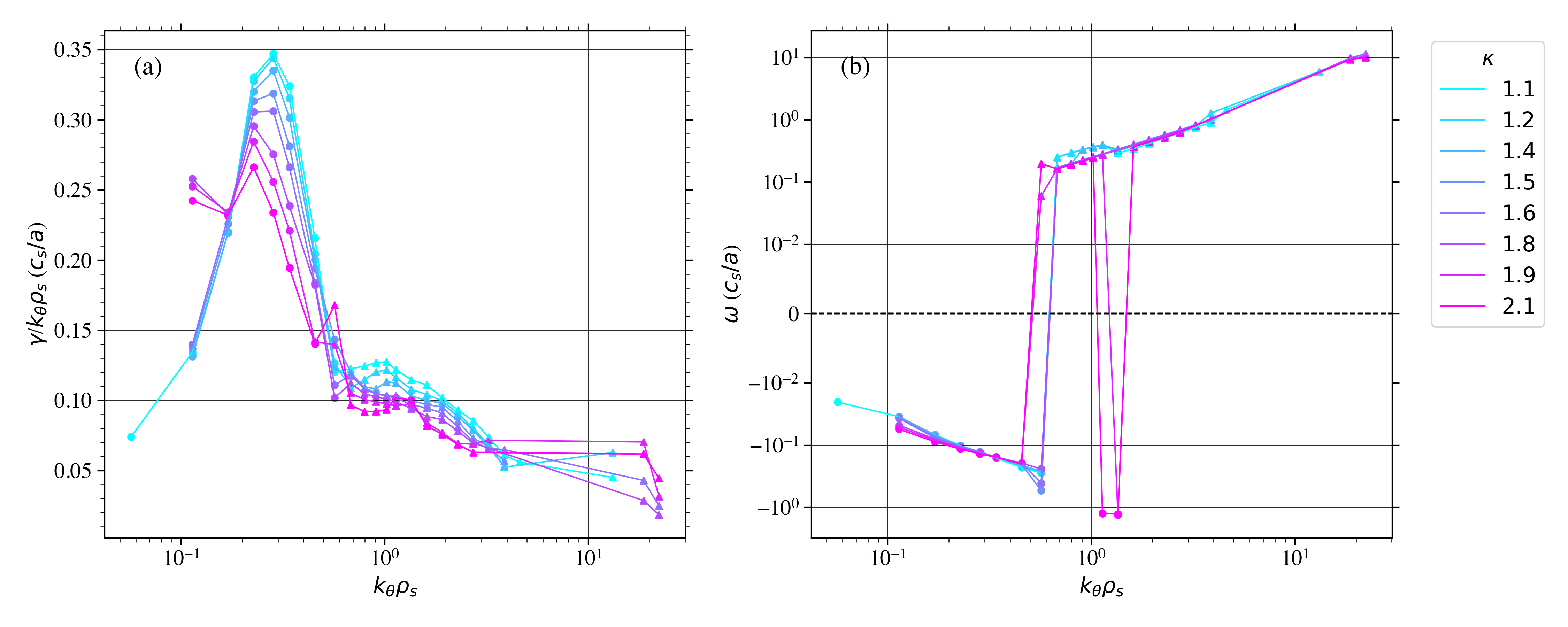}
    \end{subfigure}
    \begin{subfigure}{\linewidth}
        \center
        \includegraphics[width=0.8\linewidth]{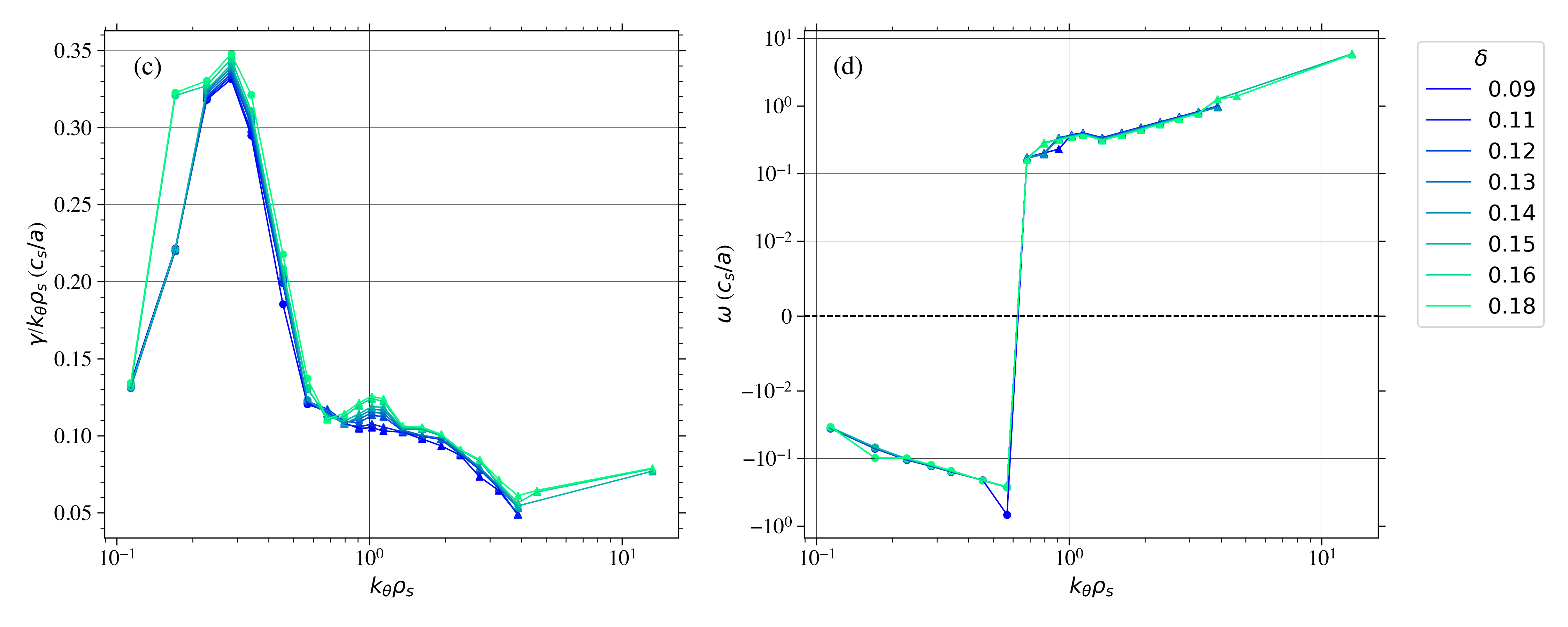}
    \end{subfigure}
    \caption{Stand-alone \texttt{TGLF} scans at $\rho = 0.6$ to assess the linear impact on turbulence of changing $\kappa$ (a, b) and $\delta$ (c,d). All of these simulations were initialized from the output of the nominal ($\kappa = 1.77$ and $\delta =0.49$ at the 99.5\% flux surface) \texttt{MAESTRO} simulation, with the composition input into \texttt{TGLF} changed externally to the \texttt{MAESTRO} framework. Both scans are shown at the nominal value for the other parameter at $\rho = 0.60$. The normalized growth rates are shown in (a) and (c), and the real frequencies, are shown in (b) and (d). Increased elongation has a strong ITG stabilization effect while increased triangularity has a weak ITG destabilization effect.}
    \label{fig:TGLF_geometry_scans}
\end{figure}

There are several key ways that plasma shape impacts fusion performance. Like impurity composition, changing core turbulent transport and pedestal pressure impact fusion performance, but now, changing plasma volume also plays a role. First, we will consider changes in the core turbulent transport. Stand-alone TGLF scans were performed in the mid-core, again at the $\rho = $ 60\% flux surface. Here, the nominal, local value of $\kappa$ is 1.37 and of $\delta$ is 0.12. The results of the $\pm 25\%$ scans are shown in Figure \ref{fig:TGLF_geometry_scans}. Increasing $\delta$ results in slightly higher ITG growth rates, meaning that higher $\delta$ linearly destabilizes the ITG, consistent with the findings of \cite{pueschelReducingTransportExtreme2024}. However, increasing $\kappa$ strongly stabilizes the ITG growth rates, linearly suppressing ITG turbulence, consistent with the findings of \cite{belliEffectsPlasmaShaping2008}.

\begin{figure}
    \centering
    \includegraphics[width=\linewidth]{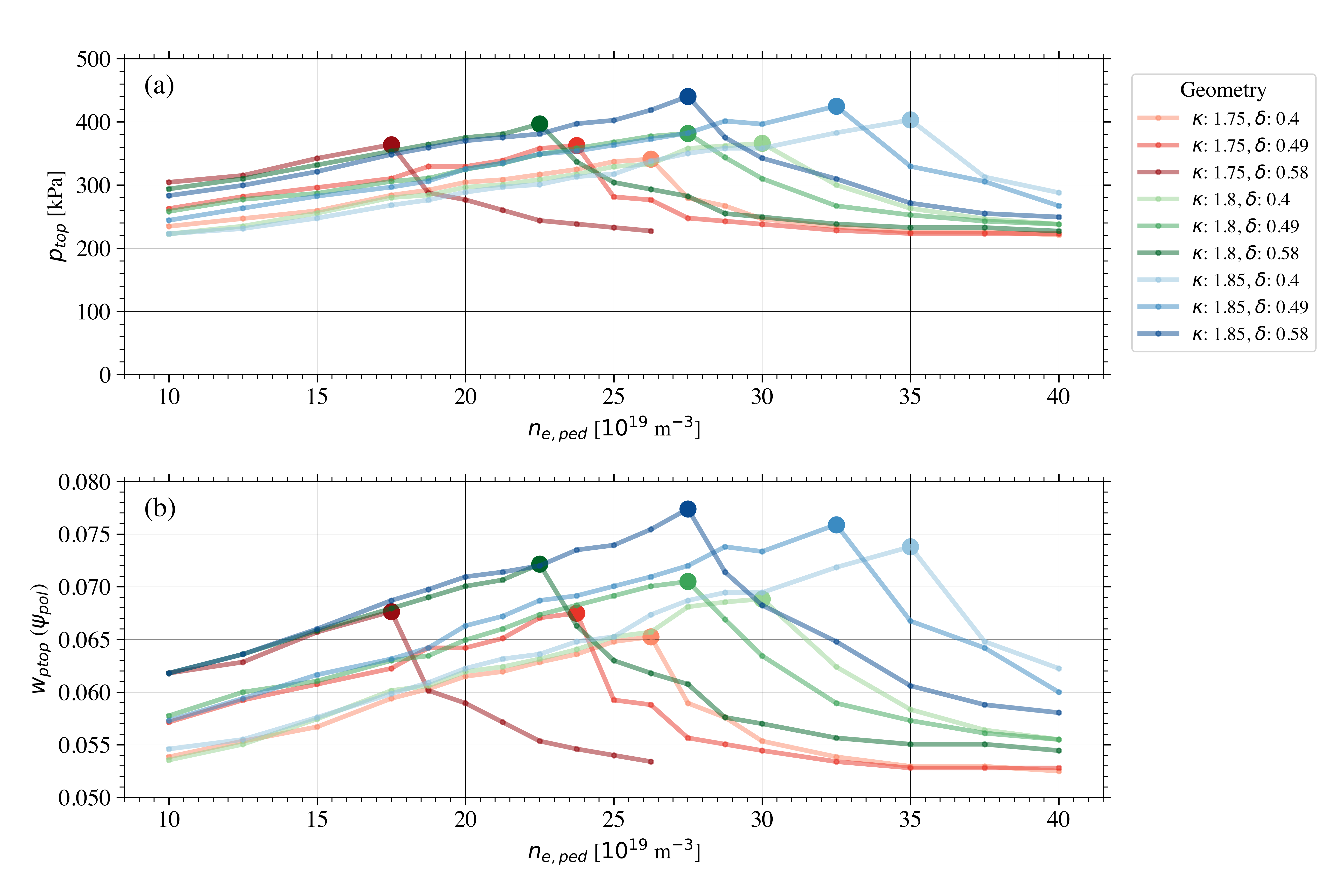}
    \caption{Impact of $\kappa$ and $\delta$ on the pedestal height ($p_{top}$) and width ($w_{ptop}$) at different pedestal densities ($n_{e,ped}$). Changes in $\kappa$ are denoted by changes in color and changes in $\delta$ are denoted by changes in shade. Increased elongation shifts the transition from a peeling-limited pedestal to a ballooning-limited pedestal to higher densities, but slightly decreases the pedestal pressure achievable for a given density in the peeling branch. Increased triangularity strongly increases the pedestal pressure achievable for a given density in the peeling branch, but slightly shifts the density at which the pedestal transitions from peeling limited to ballooning limited.}
    \label{fig:eped_kappa_delta_scan}
\end{figure}

Second, the predicted pressure at the top of the pedestal has a strong, direct dependence on both $\kappa$ and $\delta$.  As $\kappa$ increases, depicted by changing colors in Figure \ref{fig:eped_kappa_delta_scan}, the ballooning mode drive is reduced. The transition from a peeling limited pedestal to a ballooning limited pedestal shifts to much higher densities. This provides the advantage of enabling higher pedestal densities with increased maximum pressure achievable, as is seen in Figure \ref{fig:eped_kappa_delta_scan}, although Greenwald density limit still constrains the maximum possible volume averaged density. However, the pressure achievable for a given $n_{e,ped}$ on the peeling branch slightly decreases with $\kappa$. Increasing $\delta$, shown by the changing shades in Figure \ref{fig:eped_kappa_delta_scan}, stabilizes the peeling mode by reducing the trapped particle fraction and therefore bootstrap current. It causes the density at which the pedestal switches from peeling limited to ballooning limited to slightly decrease. However, the pressure achievable for a given $n_{e,ped}$ on the peeling branch increases strongly. Because the effect of $\kappa$ on the density of the switch from peeling to ballooning limited pedestals is stronger, and the effect of $\delta$ on the pressure achievable for a given $n_{e,ped}$ is stronger, highly shaped plasmas with both high $\kappa$ and high $\delta$ allow the highest pedestal pressures. 

\begin{figure}
    \centering
    \includegraphics[width=\linewidth]{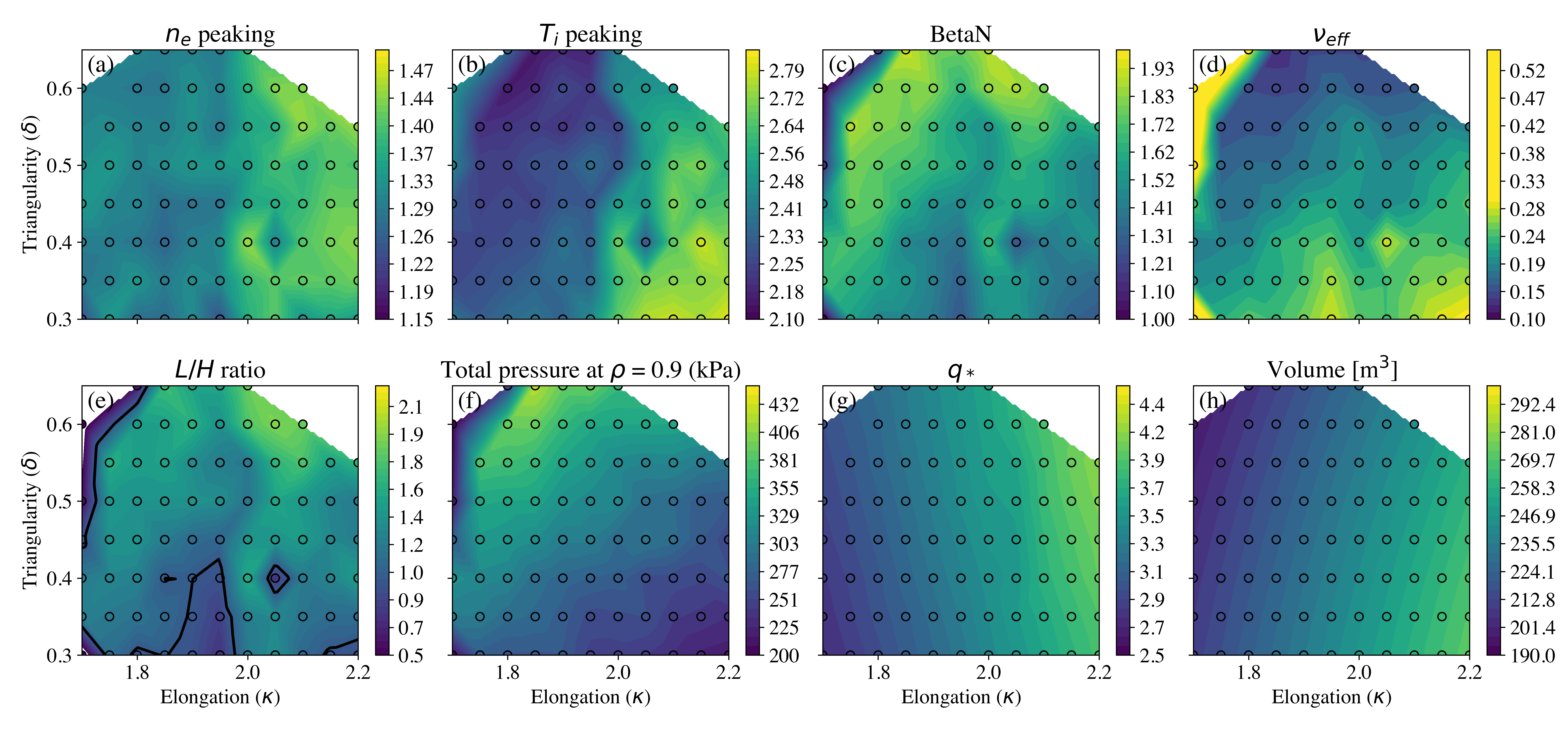}
    \caption{Additional quantities of interest for the scan of $\kappa$ and $\delta$ depicted in Figure \ref{fig:shaping_scan1}; \texttt{MAESTRO} simulations performed are circled in black. Subplots (a) and (b) show the electron density and ion temperature peaking respectively, where the peaking is the ratio between the on axis and volume averaged values. Subplot (c) shows the value of volume averaged $\beta_N$. Subplot (e) shows the effective collisionality as defined \cite{angioniScalingDensityPeaking2007} and multiplied by $Z_{eff}/2$ to relax the assumption of $Z_{eff} = 2.0$ made in that paper. Subplot (e) shows the ratio between the power threshold predicted by the scaling in \cite{martinPowerRequirementAccessing2008} and the scrape-off-layer power. Subplot (f) shows the plasma pressure of $\rho = 0.9$, which is a proxy for the pressure at the top of the pedestal as given by \texttt{EPED} (although width effects are neglected). Subplot (g) depicts the safety factor, and subplot (h) shows the plasma volume.}
    \label{fig:shaping_scan2}
\end{figure}

Lastly, the plasma volume increases strongly with increased $\kappa$ and decreases weakly with increased $\delta$, as can been seen in Figure \ref{fig:shaping_scan2}h. The core turbulent transport and pedestal pressure effects impact the fusion power density, but absolute fusion power is also dependent linearly on volume. Therefore, through volumetric effects, increasing $\kappa$ substantially benefits the total fusion power, and increasing $\delta$ slightly reduces it. 

Now, the combined impact of changing plasma shape, via the three mechanisms discussed, will be examined. ITG stabilization, as seen in the stand alone \texttt{TGLF} simulations, shows up in the resultant increase in temperature peaking, shown in Figure \ref{fig:shaping_scan2}b. Density peaking, shown in Figure \ref{fig:shaping_scan2}a also increases with higher values of elongation. Given this scan is performed at fixed density, firmly on the peeling branch, increased pedestal pressure is found at low elongation and high triangularity, as can be seen in Figure \ref{fig:shaping_scan2}f. Figure \ref{fig:kappa_scan_profiles}h, found in the appendix, shows the impact of changing elongation on \texttt{EPED-NN} scans of $\kappa$ and $n_{e,ped}$. To access the higher maximum pedestal pressure feasible with increased elongation, density would have to increase. Again, the majority of the parameter space is above the L-H transition threshold, seen in Figure \ref{fig:shaping_scan2}e, with a notable amount of margin. Assuming access to a pedestal in the performance predictions is reasonable, especially when elongation and triangularity are sufficiently large. Since this scan is performed at a fixed total plasma current of 12 MA, the safety factor, $\text{q}^*$ increases with increasing elongation and slightly decreases with increasing triangularity. While higher elongation increases the difficulty of vertical stability control, the lower values of $\text{q}^*$ suggest potentially reduced disruptivity. Alternatively, we could have elected to assume a fixed value of $\text{q}^*$ and modified the plasma current accordingly. In this scenario, plasma current would be able to increase with increasing $\kappa$, thereby increasing the pedestal pressure. In work not depicted here, it was shown fusion power would increase even more strongly with $\kappa$ if $\text{q}^*$ were held fixed instead of plasma current. However, increased plasma current would lead to shorter plasma duration and larger disruptive forces.

We now turn our attention to the next higher order parameter of squareness ($\zeta$), which is here defined consistent with MXH shaping definitions. This work is enabled by access to a version of \texttt{EPED} that takes as input the MXH sine coefficients, not just $\delta$. By enforcing a value of zero for the MXH cosine coefficients, up-down symmetry is enforced. We perform this scan near the optimal elongation, at $\kappa = 2.1$, and the nominal plasma composition.  The range of squareness from -0.2 to 0.2 is scanned. The nominal value from the equilibrium provided in \cite{hillesheimOverviewPhysicsBasis2026} gives a negative squareness. However, optimization for squareness was not yet performed on this design, and the nominal value used for performance predictions is zero squareness. 

\begin{figure}
    \centering
    \includegraphics[width=0.7\linewidth]{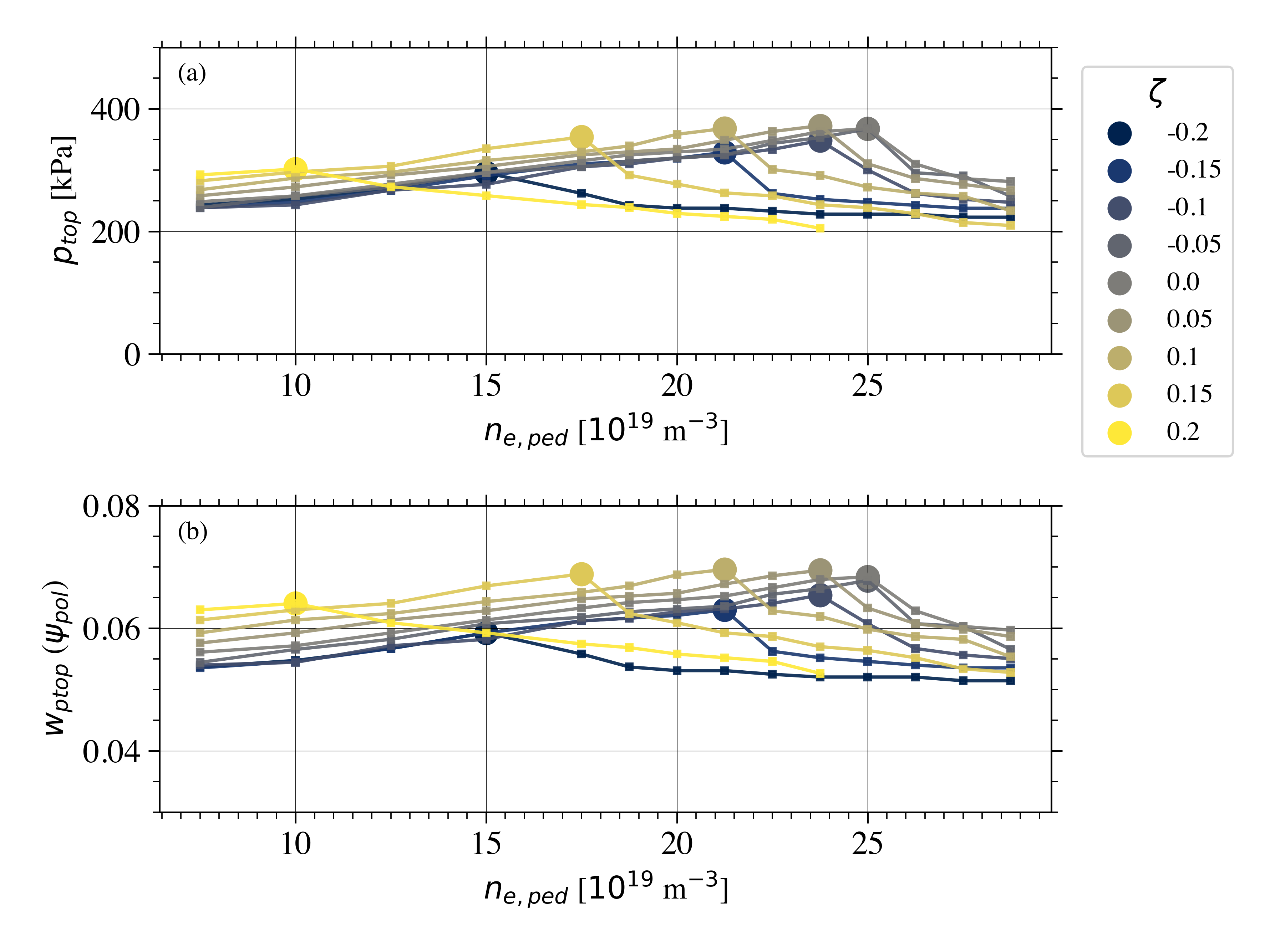}
    \caption{Impact of squareness ($\zeta$) on the pedestal height ($p_{top}$) and width ($w_{ptop}$) at different pedestal densities ($n_{e,ped}$) for $\kappa = 1.77$ and $\delta = 0.49$.  For an ARC-like device, increasing absolute value of $\zeta$ decreases the density at which the pedestal transitions from peeling limited to ballooning limited. Like triangularity, increasing $\zeta$ increases the pressure achievable at a given density on the peeling branch.}
    \label{fig:eped_zeta_scan}
\end{figure}

First, a stand-alone \texttt{EPED} scan of squareness is performed, as shown in Figure \ref{fig:eped_zeta_scan}. The highest pedestal pressure can be achieved for zero squareness. As the absolute value of the squareness increases, the density at which the pedestal changes from peeling limited to ballooning limited decreases. However, like $\delta$, the pressure achievable for a given $n_{e,ped}$ on the peeling branch increases with increasing squareness. Additionally, a stand alone \texttt{TGLF} scan was performed, not shown in this paper, which found negligible linear ITG stabilization resulting from increased squareness. 

\begin{figure}[h]
    \centering
    \begin{subfigure}{0.49\linewidth}
        \includegraphics[width=\linewidth]{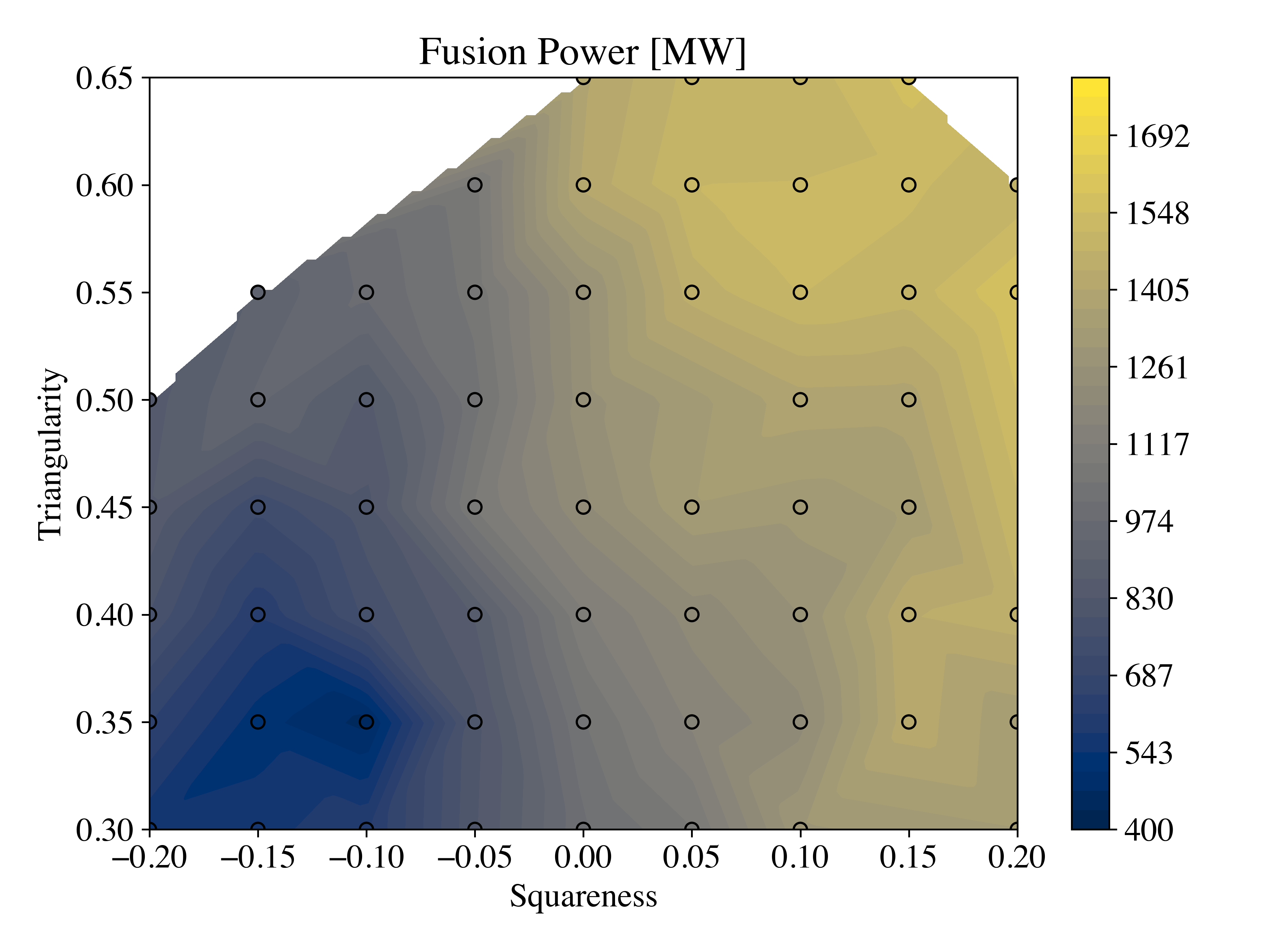}
    \end{subfigure}
    \begin{subfigure}{0.49\linewidth}
        \includegraphics[width=\linewidth]{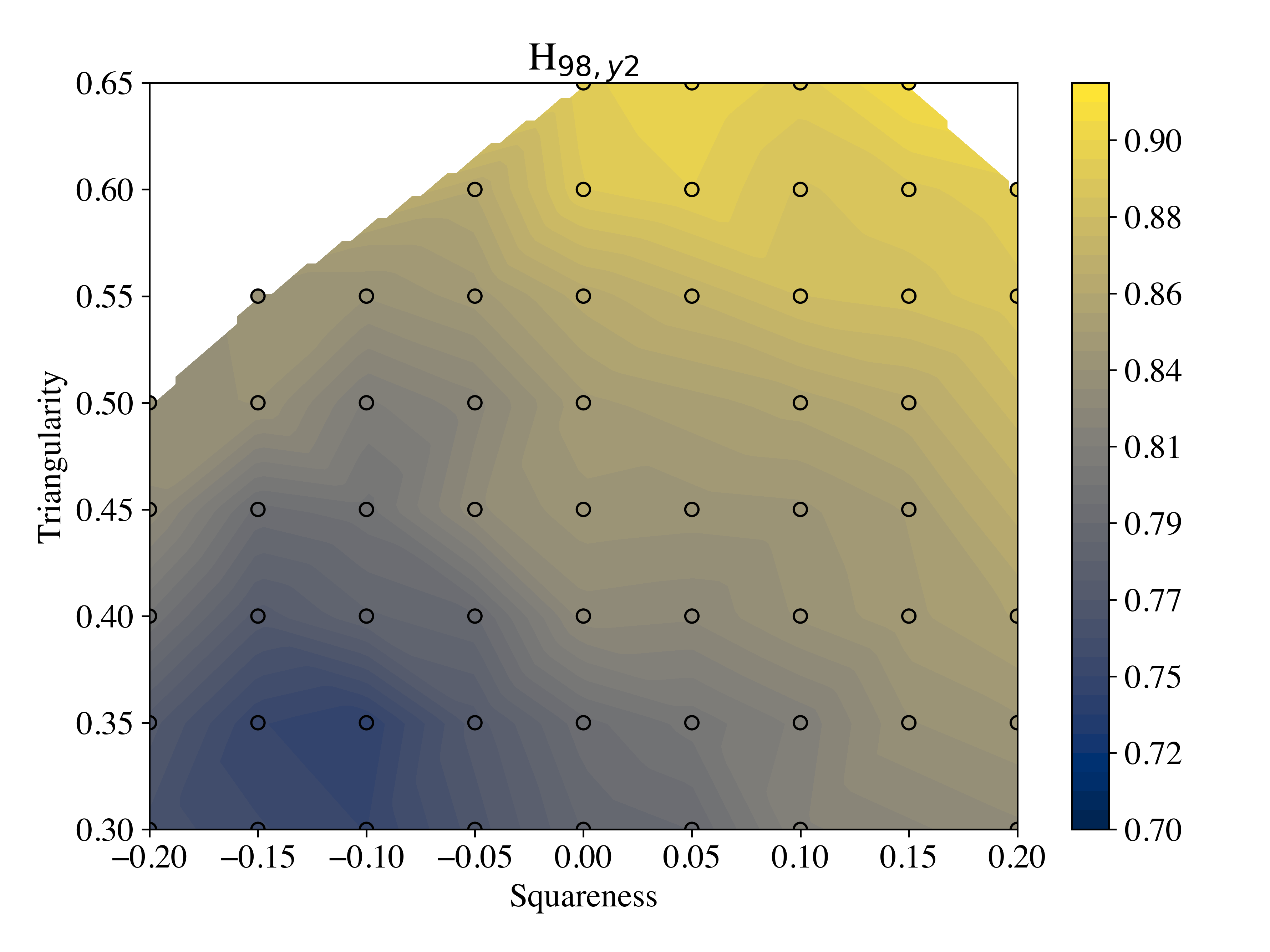}
    \end{subfigure}
    \caption{The resulting (a) fusion power and (b) scalar multiplicative factor on the energy confinement time predicted by the ITER98(y,2) scaling \cite{iterphysicsexpertgrouponconfinementandtransportChapter2Plasma1999} from performing a two dimensional scan of squareness ($\zeta$) and triangularity ($\delta$) at $\kappa = 2.1$. Nominally, $\zeta = 0$ and $\delta = 0.49$ are assumed.
    The circles show the locations of \texttt{MAESTRO} simulations performed with interpolation between them. By going to larger, positive squareness higher fusion power and better confinement can be achieved.}
    \label{fig:squareness_scan1}
\end{figure}

Then, a brute force scan of triangularity and squareness was performed. Both fusion power, shown in Figure \ref{fig:squareness_scan1}a, and confinement, shown in Figure \ref{fig:squareness_scan1}b, are enhanced by higher triangularity and squareness. This confirms if the triangularity is limited by engineering constraints, squareness could potentially be leveraged to compensate. An additional benefit of positive squareness is that volume increases with squareness. The favorable effect seen in this 2D scan of high squareness requires the density be such the pedestal is firmly peeling limited. If this scan were conducted at a lower $\kappa$, lower squareness might be preferable to prevent ballooning modes from taking over. That elongation, triangularity, and squareness, as well as $Z_{eff}$ and $f_{main}$, have different impacts on the same three key levers, core turbulent transport, pedestal pressure, and volume, suggests the importance of the joint optimization performed in the next section. 


\section{Bayesian Optimization Results} 
\label{sec:BO}

To do higher dimensional scans, use of Bayesian optimization is necessitated by the computational cost of each \texttt{MAESTRO} simulation,  $\sim$100 cpu hours when run on 16 cores of a AMD EPYC cluster-based node. Bayesian optimization is a technique that allows the efficient optimization of a computationally expensive black-box function (a function for which an analytical description is not available). We encourage the reader to see \cite{frazierTutorialBayesianOptimization2018} for a comprehensive overview of Bayesian optimization, but a brief description is provided here. There are three key steps in the Bayesian optimization process: (1) evaluating the black box function, (2) training the surrogates on all past evaluations of the black box function, and (3) using the acquisition function to determine the next set of inputs that would be most valuable to evaluate with the black box function.  For each optimization, we first produce a set of training data by using Latin Hypercube sampling, which efficiently fills the space \cite{mckayComparisonThreeMethods1979}. A sufficient number of training points are needed to map the space sufficiently well to accelerate the finding of the global optimum. Surrogate functions are fit to the training data to make an estimated, continuous mapping of the space. The surrogates used here are Gaussian processes \cite{gramacySurrogatesGaussianProcess2021} with covariances assumed to be described by the Mat\'ern kernel \cite{Matrn1960SpatialV}. The acquisition function quantifies the trade off between increasing the amount of information gained by the next evaluation, by sampling the set of inputs for which there is the most uncertainty in the outcome, and preferring sets of inputs that are most likely to approach the optimum. Here, we use a Monte Carlo logarithmic expected improvement acquisition function \cite{NEURIPS2023_419f72cb} with batches of 10 evaluations performed in parallel \cite{BatchingBoTorch2026}. The BoTorch \cite{balandatBoTorchFrameworkEfficient2020} package was used for optimization. 

 The boundary of allowed input parameters for the optimization is shown in Table \ref{tab:variable-values}; these were selected so that they did not limit the optimal solution, with the exception of enforcing a maximum possible shaping. As discussed in the previous section, the elongation at the 99.5\% flux surface was limited to $\kappa < 2.0$ due to vertical stability considerations, the triangularity was limited to $\delta < 0.65$ for feasibility with reasonable shaping coils, and the squareness was limited to $-0.20 < \zeta < 0.20$. Three optimizations were performed. In the first, all six parameters were allowed to vary over the specified ranges. In the second, the elongation is fixed to the nominal value determined feasible in the ARC V3A design. In the third, since squareness is a more difficult, less frequently controlled parameter experimental, optimization was performed enforcing $\zeta = 0$.

For the optimization objective the fusion power was selected, with penalties for the volume averaged density exceeding the Greenwald density limit ($f_G > 0.9$) or failing to have the predicted scrape-off-layer power meet the Martin L-H transition threshold power ($f_{LH} < 1$) \cite{martinPowerRequirementAccessing2008}. The conservative choices of enforcing a limit on the volume averaged density and penalizing solutions greater than 90\% of the Greenwald density is for consistency with the ARC V3A design \cite{hillesheimOverviewPhysicsBasis2026}.The formulation for the objective, which we seek to maximize, is: 
\begin{equation}
    \text{Objective} = \text{success probability} \times P_{fus} \times min(\frac{1}{[f_G + (1-0.9)]^5}, 1) \times max(f_{LH}, 1)
\end{equation} 
An additional term, success probability, is zero if the \texttt{MAESTRO} run does not converge, which can occur if an unfeasible geometry or impurity is chosen, or if poor performance leading to no possible steady-state solution. While this value is binary for \texttt{MAESTRO} evaluations, the surrogate can predict fractional probabilities of success, enabling evaluation time to be spent on cases more likely to produce useful results. \

\begin{table}[h]
    \centering
    \caption{Optimization boundaries, optimal values of inputs, and resulting performance. Elongation ($\kappa$), triangularity ($\delta$), and squareness ($\zeta$) are specified at the 99.5\% flux surface.}
    \begin{tabular}{lcccccc}
        \toprule
        Variable & Minimum & Maximum & Nominal & 6D & 5D (fixed $\kappa$) & 5D (fixed $\zeta$) \\
        \midrule
        $Z_{eff}$ & 1.5 & 2.5 & 1.5 & 2.1 & 1.70 & 2.4  \\
        $f_{main}$ & 0.70 & 0.90 & 0.85 & 0.88 & 0.83 & 0.86\\
        $\kappa$ & 1.70 & 2.00 & 1.77 & 1.99 & 1.77 & 1.92 \\
        $\delta$ & 0.45 & 0.65 & 0.49 & 0.60 & 0.50 & 0.60 \\
        $\zeta$ & -0.20 & 0.20 & 0 & 0.17 & 0.12 & 0 \\
        $n_{e,ped} [\times 10^{20}/m^3]$  & 1.9 & 2.5 & 2.10 & 2.18 & 2.03 & 2.20 \\
        \cmidrule{1-7}
        $P_{fus} [MW]$ & & & 1140 & 1910 & 1537 & 1515 \\
        $\tau_{flattop}$ [s] & & & 1009 & 1043 & 1208 & 842 \\
        
        \bottomrule
    \end{tabular}
    \label{tab:variable-values}
\end{table}

\begin{figure}
    \centering
    \includegraphics[width=\linewidth]{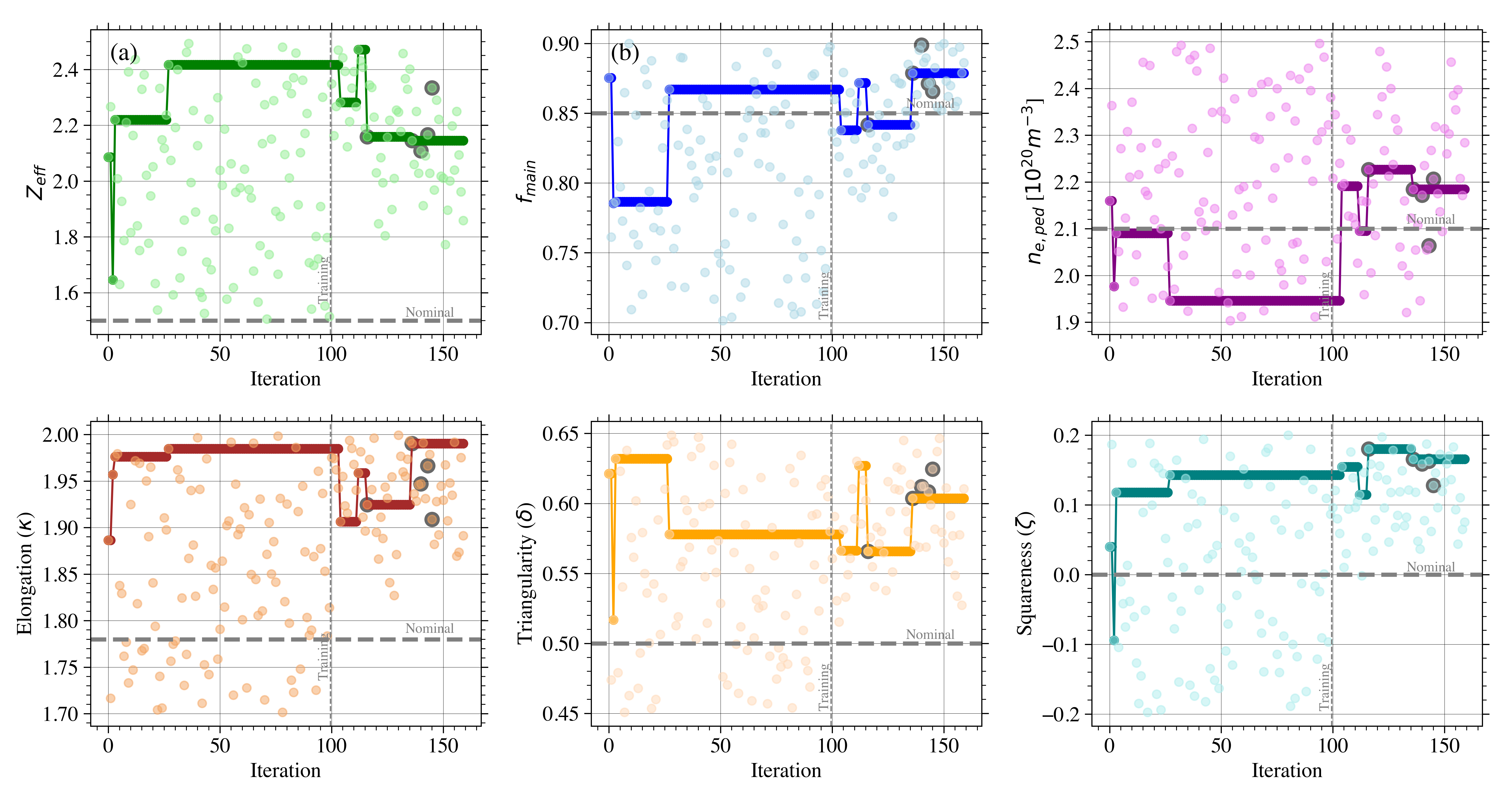}
    \caption{Input parameter values for each iteration of the full 6D core, steady state operational optimization. Grey circles denote the iteration achieved greater than 1800 MW of fusion power The iteration thus far with the best residual is highlighted. The nominal value of each input parameter for ARC V3A is shown with a horizontal dashed line. Evaluations to the left of the vertical dashed line labeled ``training" are selected via Latin hypercube sampling, while evaluations to the right are selected by the acquisition function. The corresponding model outputs can be found in Figure \ref{fig:6D_Pfus_by_iteration}.}
    \label{fig:6D_Best_params_by_iteration}
\end{figure}

The evolution of the optimal parameters, for the full 6D optimization, as a function of iteration is shown in Figure \ref{fig:6D_Best_params_by_iteration}, with the input parameters of the best evaluation shown in Table \ref{tab:variable-values}. In Figure \ref{fig:6D_Best_params_by_iteration}, grey dots highlight all evaluations that exceeded 1800 MW of fusion power, allowing for the range of inputs affording optimal solutions to be seen. A higher optimal $Z_{eff}$ than the nominal ARC V3A design, Zeff = 2.1, is found, with all near optimal solutions at substantially elevated values with respect to the nominal design. A higher main ion fraction, $f_{main}$ = 0.88, is preferred. However, near optimal solutions span values both above and below the nominal. An elongation near the maximum considered is best, $\kappa = 1.99$, although lower values can produce near optimal solutions. An elevated ideal triangularity of $\delta = 0.60$ is found; although, a relatively wide, elevated range of values can give the near optimal solutions. A large positive squareness is found to improve performance, with the optima occurring at $\zeta = 0.17$. These results are all in line with the scans done in the previous section. Finally, slightly higher values of $n_{e,ped}$ than selected for ARC V3A are found to be most preferable, although there is again a range of values for near favorable solutions. 

\begin{figure}
    \centering
    \includegraphics[width=\linewidth]{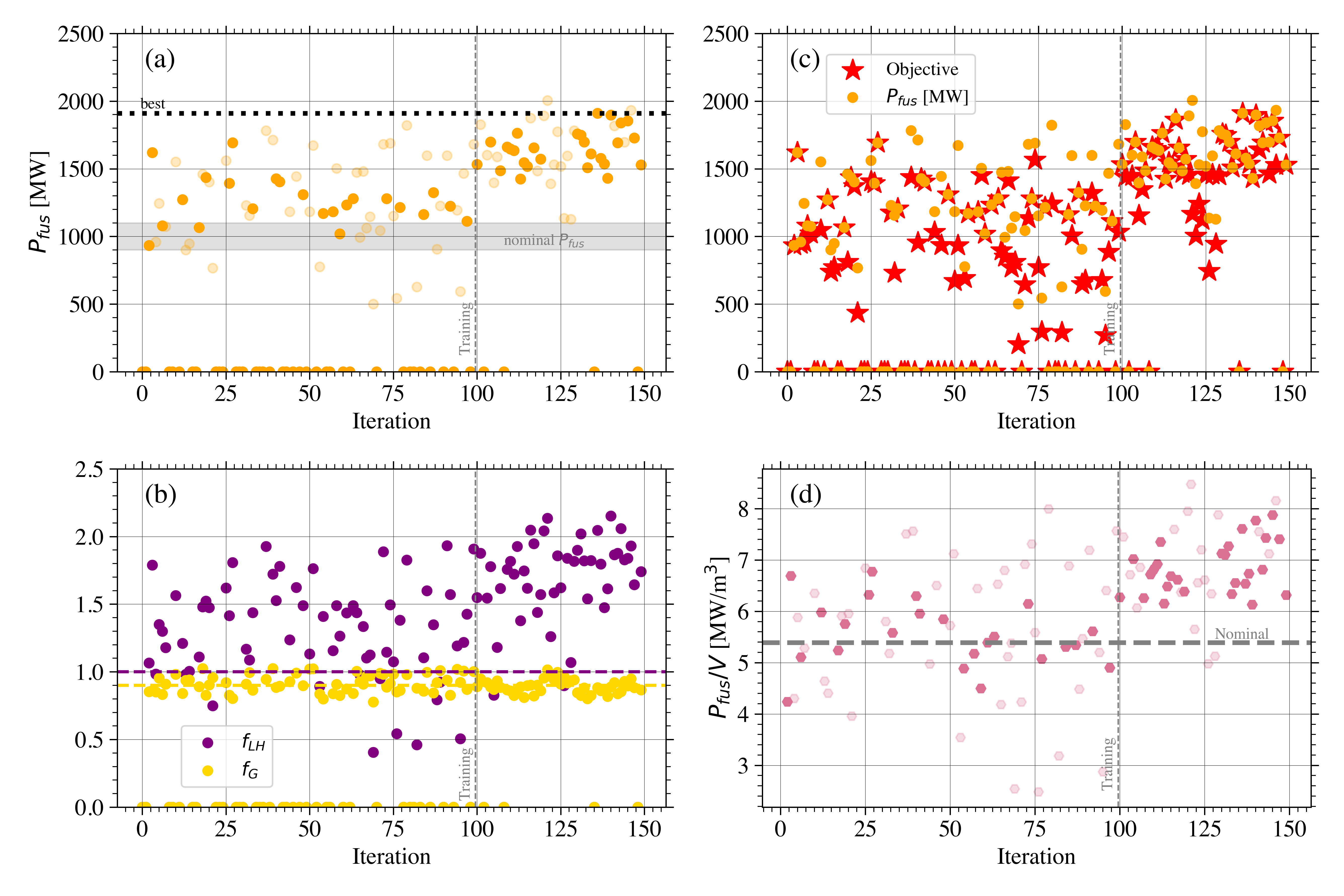}
    \caption{Results of full 6D core, steady state operational optimization. Subplot (a) shows the fusion power, which increases by over 65\%, to a maximum of 1910 MW through optimization. The nominal fusion performance of ARC V3A as found in \cite{howardPerformanceTransportARC2026} is shown in the shaded region. The iteration with the best objective is shown with the black dotted horizontal line. Iterations with $f_{LH} > 1$ and $f_G < 0.9$ are fully saturated while iterations for which either of these conditions is violated are in transparent orange.
    The value of $f_G$ and $f_{LH}$ are shown in subplot (b), with the dotted lines showing the respective thresholds. 
    Subplot (c) shows the objective value. Since the objective is the fusion power with a penalty for exceeding the Greenwald density limit ($f_G>0.9$) or not reaching the L-H transition power threshold ($f_{LH}<1$), iterations where the fusion power dot is not in center of the objective star denote those where a penalty was applied.
    Subplot (d) shows the volumetric fusion density with the nominal value for ARC V3A shown in the grey dashed line.
    Evaluations to the left of the vertical dashed line labeled ``training" are selected via Latin hypercube sampling, while evaluations to the right are selected by the acquisition function. The corresponding model inputs can be found in Figure \ref{fig:6D_Best_params_by_iteration}.}
    \label{fig:6D_Pfus_by_iteration}
\end{figure}

The best set of input parameters results in a fusion power of 1910 MW, more than 65\% higher than the fusion power predicted for ARC V3A, assuming no squareness, with the same modeling assumptions. Fusion power versus iteration can be seen in Figure \ref{fig:6D_Pfus_by_iteration}a. Iterations for which both the constraints ($f_G<0.9$ and $f_{LH}>1$) are met, are highlighted in this figure. The constraint values versus iteration are seen in Figure \ref{fig:6D_Pfus_by_iteration}b. Iterations with substantial fusion power easily reach the $f_{LH} > 1$ criterion. However, the density limit constraint is less readily satisfied, even though there is a quintic penalty in the objective fusion for exceeding it. This is in contrast to the findings in \cite{slendebroekExploringFusionPower2026}, likely because of the conservative scape-off-layer power enforced in that work. The values of the objective function can be seen in Figure \ref{fig:6D_Pfus_by_iteration}c. Recall, when both constraints are met, the objective function is the same value as the predicted fusion power. \textcolor{black}{The pulse duration, as calculated by \cite{bodyCfsenergyCfspopconV7022024, barrPowerbalanceModelLocal2018}, is shown in Table \ref{tab:variable-values}. No reduction in flattop duration is expected in the adjust scenario. While increased $Z_{eff}$ reduces the flattop time, this is countered by an increase in elongation and temperature.} Finally, the volumetric fusion power density is depicted in Figure \ref{fig:6D_Pfus_by_iteration}d. The nominal value for ARC V3A predicted by this workflow is $5.39 \text{MW}/\text{m}^3$. Fusion power densities above $7.75 \text{MW}/\text{m}^3$ are achieved meeting both constraints, with 6D optimization. In other words, assuming a constant volumetric fusion power density, the same fusion power as the nominal ARC V3A design could be achieved with 30\% less plasma volume\textcolor{black}{, suggesting the potential feasibility of cost reductions in future design iterations}.

\textcolor{black}{For the second optimization, an elongation equal to that of ARC V3A was enforced since increased elongation makes vertical stability more challenging. An operating point, shown in Table \ref{tab:variable-values}, was found with with this constraint that produces 1537 MW of fusion power. This is more than $30\%$ higher than the nominal ARC V3A performance.} The third optimization, enforcing zero squareness, was performed should achieving strong positive squareness prove difficult for ARC-like devices or insufficiently robust to perturbations. The best values of the input parameters for this optimization can be found in Table \ref{tab:variable-values}, which are similar to the results from the 6D optimization. A fusion power of 1515 MW was found for these inputs, which is also over 30\% better than the nominal ARC V3A performance. The inclusion of squareness allows a 25\% higher fusion power than the fixed squareness optimization. 

\section{Discussion} 
\label{sec:discussion}

In this work, we have performed an optimization \textcolor{black}{near the design space of ARC V3A} that considers the potential impact of changing impurity composition ($Z_{eff}$ and $f_{main}$), plasma shape ($\kappa$, $\delta$, $\zeta$) and pedestal density ($n_{e,ped}$). The modeling was performed with the \texttt{MAESTRO} framework, which couples together physics based models for predicting the equilibrium, heating, current diffusion, turbulent transport, and pedestal pressure. We observe the importance of this optimization being performed with a 1.5D physics-based model. Commonly used tools in reactor design are 0D models, which rely on empirical scaling laws to predict the energy confinement time. The temperature and density of the plasma are also assumed. In Figures \ref{fig:comp_scan}b, \ref{fig:shaping_scan1}b, and \ref{fig:squareness_scan1}b, the ad-hoc factor one would have to multiply the results of the ITER98(y,2) scaling law by to recreate the confinement time predicted by the physics-based simulations is shown. Here, we find non-negligible changes in confinement resulting from changing impurity composition and plasma shape that are not captured by the scaling law. While $\kappa$ is included in the scaling law, its dependence is actually too strong. This precludes the selection of an optimal, realistic, operating point with empirically-based modeling. Often, once a design is selected, physics-based 1.5D modeling is employed. However, key design variables, like device size, aspect ratio, plasma current, and magnetic field strength all have strong effects on the pedestal pressure and turbulent transport. Therefore, we believe it is critical to employ physics-based operational optimization in parallel with the design process. One example of a common effect when this optimization is not done is shown in the appendix in Figure \ref{fig:kappa_scan_profiles}h. Here, $n_{e,ped}$ is not optimized with changing $\kappa$. The benefit of shifting the density at which ballooning modes limit the pedestal to higher values is not realized when $n_{e,ped}$ is held fixed. Similar behavior would be seen when changing design variables. This work presents a workflow for fast operational optimization that could be performed at the design stage, with a first application of it to the ARC V3A design \cite{hillesheimOverviewPhysicsBasis2026}.

For changes in $Z_{eff}$ and $f_{main}$, fusion performance is impacted by (1) changing the density of the main, fuel ions, (2) changing the turbulence via ITG stabilization, (3) changing the radiative sink term in the steady state power balance, (4) changing the pressure of the pedestal. At higher $Z_{eff}$, reducing $f_{main}$ provides more ITG stabilization than at lower $Z_{eff}$, consistent with results found in \cite{rodriguez-fernandezStudiesImpurityinducedTurbulence2025}. The very weak dependence of fusion power on $f_{main}$, seen in Figure \ref{fig:comp_scan}a, is likely due to the negative effect of fuel dilution and positive effect of ITG stabilization canceling out. Higher $Z_{eff}$ than is nominally assumed in ARC V3A is favorable. By optimization of just impurity composition, at fixed $n_{e,ped}$, fusion power can be increased by 25\% over the nominal. Allowable increased core $Z_{eff}$ and decreased core $f_{main}$ is favorable to the difficult challenge of finding a core-edge integrated solution. 

 Changing plasma shape also impacts fusion performance through (1) changing the core turbulent transport and (2) changing the pedestal top pressure. Additionally, changing shape (3) changes the plasma volume. The turbulent transport is stabilized by increases in $\kappa$ and decreases in $\delta$. As a higher order shaping parameter, $\zeta$ has a limited effect on core turbulent transport. The pedestal pressure at a fixed $n_{e,ped}$ on the peeling branch increases with $\delta$ and $\zeta$ but slightly decreases with $\kappa$. However, the density at which ballooning modes limit the pedestal increases with $\kappa$ but slightly decreases with $\delta$. For this design, non-zero $\zeta$ causes the density at which ballooning modes limit the pedestal to decrease. Therefore, for high $\zeta$ to be advantageous, sufficiently high $\kappa$ must be employed. Both increased $\kappa$ and $\zeta$ increase the plasma volume while increased $\delta$ reduces the plasma volume. The distribution of the plasma within the volume can also change the total fusion power, which is the idea that underlies volumetric optimization \cite{parisiDoublingFusionPower2025}. While this effect is captured by this workflow, we do not specifically explore its consequence here.  Putting all of these effects together, relatively high values of $\kappa$, $\delta$, and $\zeta$ are all preferred. \textcolor{black}{However, the authors recognize the potential difficulty of achieving a design consistent with such strongly shaped operation.} By optimization of just $\kappa$ and $\delta$, at fixed $n_{e,ped}$, for $\kappa > 2.0$, fusion power can be increased by 20\% over the nominal. A $\sim$10\% increase in fusion power, over zero squareness, can be achieved by going to higher positive squareness. 
 
 To enable optimization in 6D operational space, efficient Bayesian optimization techniques are leveraged instead of the expensive brute force scan technique used to build intuition in the later sections of the paper. A maximum fusion power of \textcolor{black}{1910} MW is attained, which represents a \textcolor{black}{65\%} increase over the nominal performance predicted by this workflow for ARC V3A. \textcolor{black}{Holding either squareness or elongation to the nominal ARC V3A value still enables a 30\% increase in fusion power to be achieved with optimization.} Higher fusion power densities than are predicted for the nominal ARC V3A design are found, suggesting the possibility of reducing the plasma volume and device cost, while maintaining the same fusion power as the nominal ARC V3A design. The trends for the input parameters are the same as those found for the 2D scans; higher $Z_{eff}$, $\kappa$, $\delta$, and $\zeta$ than the nominal design are preferred. Also like the 2D scans, there is limited dependence found on the value of $f_{main}$. \textcolor{black}{Pulse duration is not expected to be negatively impacted by this optimization.} It is important to recognize there are heuristics in this optimization process, and there is no guarantee than an even more favorable solution with different input parameters does not exist. However, as performance predictions are done with the full \texttt{MAESTRO} model (not a surrogate of it), it is certain that \texttt{MAESTRO} will predict an elevated value of fusion performance, with respect to the nominal ARC V3A design, for these inputs.
 Since a value of $n_{e,ped}$ that puts the density at the Greenwald density limit is ideal, the density of the transition from a peeling-limited pedestal to a ballooning-limited pedestal is less relevant, and there is reduced coupling between the the impurity composition and geometric shaping optimizations in this case.  Because density is such a strong lever on fusion power, if the higher densities predicted for devices like ARC by recent works including a power dependence in the density limit prove feasible \cite{giacominFirstPrinciplesDensityLimit2022, manzPowerDependenceMaximum2023, marisCorrelationLmodeDensity2024a, marisPredictionControlTokamak2026}, the maximum achievable fusion power could increase substantially. Exploration of the impact of different density limits would be a very interesting avenue for future work.

 Appropriate uncertainty quantification for this analysis can be found in \cite{howardPerformanceTransportARC2026}. Sources of uncertainty considered in that work include changes to the pedestal pressure by \texttt{EPED}, the ratio of ion temperature to electron temperature at the top of the pedestal ($T_{i,top}/T_{e,top}$), the ratio of the separatrix density to the pedestal density ($n_{sep}/n_{ped}$), and the tungsten concentration, as depicted in Figure 5 of \cite{howardPerformanceTransportARC2026}. The impact of \texttt{TGLF} modeling choices, a rough proxy for uncertainty in the rate of turbulent transport, is shown in Figure 11 of \cite{howardPerformanceTransportARC2026}. According to this work, the main drivers of the uncertainty in the fusion power predictions is the uncertainty in the pedestal pressure and the uncertainty in the core turbulent transport. Validation studies of \texttt{EPED} have found it agrees with experimental results to about 20\% \cite{snyderFirstprinciplesPredictiveModel2011, walkCharacterizationPedestalAlcator2012}, which results in a 30\% uncertainty in the fusion power predicted by \texttt{MAESTRO} in ARC V3A \cite{howardPerformanceTransportARC2026}. Consideration of the variation in fusion power predicted by the saturation rules newer than \texttt{SAT1geo} \cite{staeblerGeometryDependenceFluctuation2021, staeblerVerificationQuasilinearModel2021, duddingNewQuasilinearSaturation2022} suggest an uncertainty of 15\% \cite{howardPerformanceTransportARC2026}. Assuming the errors are independent and multiplicative, adding them in quadrature gives us a relative uncertainty on fusion power of 34\%, which should be considered applicable to fusion performance predictions made in this paper. 

 While one key limitation of this work is the absence of divertor modeling, extensive boundary modeling was performed in \cite{eichPowerParticleExhaust2026}, which concluded ARC V3A had a reasonable path towards edge integration. Increasing $Z_{eff}$ and decreasing $f_{main}$, as suggested by this work, would only be more favorable. Another limitation of this work is that impurity transport channels are not modeled. A fixed $Z_{eff}$ profile throughout the core is assumed. This means that impurity modes, which are dependent on the shape of the impurity density profile with respect to the main ion density profile, are not self consistently modeled. Recent work has shown that a flat $Z_{eff}$ profile approximation works reasonably well for SPARC \cite{muracaImpurityPeakingSPARC2026} and ARC \cite{muracaCoreedgeIntegratedModeling2026}. Another limitation of this modeling, particularly with respect to shaping scans, is the generation of the equilibrium and initial safety factor profile. The equilibria here are generated using the MXH formulation from the geometry desired at the 99.5\% flux surface but the exact equilibria in the outer flux surfaces, including feasible coil placements, remains an area of future work. The results of 1.5D physics models, including \texttt{MAESTRO} are sensitive to the initial safety factor profiles. There is also sensitivity to the initial $\beta_N$ condition assumed because of the $\beta_N$ dependence of \texttt{EPED}. Further characterization of the sensitivity to initial conditions, as well as full time-dependent modeling, is important future work.

Other future work includes inclusion of plasma current as a variable. Inclusion of a core-edge integration model, like \cite{bodySimpleAccurateModel2025}, would help ensure proposed core solutions are viable at the boundary.  Trajectory optimization, would be a valuable additional step ensure there is a pathway to achieving the proposed steady-state solutions.  Higher fidelity modeling, with gyrokinetic models and/or impurity transport, to confirm the results found here, would also be valuable. Design optimization, particularly modifications to the aspect ratio and magnetic field strength on axis, is of interest. The range of parameters available for operational optimization, especially the plasma shaping, are determined by the tokamak design, including the vacuum vessel shape and coil placements. Sensitivity analyses would include the impact of the pedestal pressure and density limit uncertainties on the optimal operating point.  As operational optimization allows for performance to be improved for an as built device, leaving room in the design phase for adjustments during the operation phase could be a valuable way to mitigate the risk from these uncertainties.

\section{Acknowledgments}
This work is supported by Commonwealth Fusion Systems, under RPP020. ChatGPT, Claude, and CoPilot assisted in code generation. The authors acknowledge the MIT Office of Research Computing and Data for providing high performance computing resources that have contributed to the research results reported within this paper. MITIM-fusion version \texttt{90322e0c} and GACODE version \texttt{0e6c00ed3} were used.

\appendix
\section{Appendix A}

Reference profiles from the 2D scans shown in section \ref{sec:scans} are provided here.

\begin{figure}
    \centering
    \includegraphics[width=\linewidth]{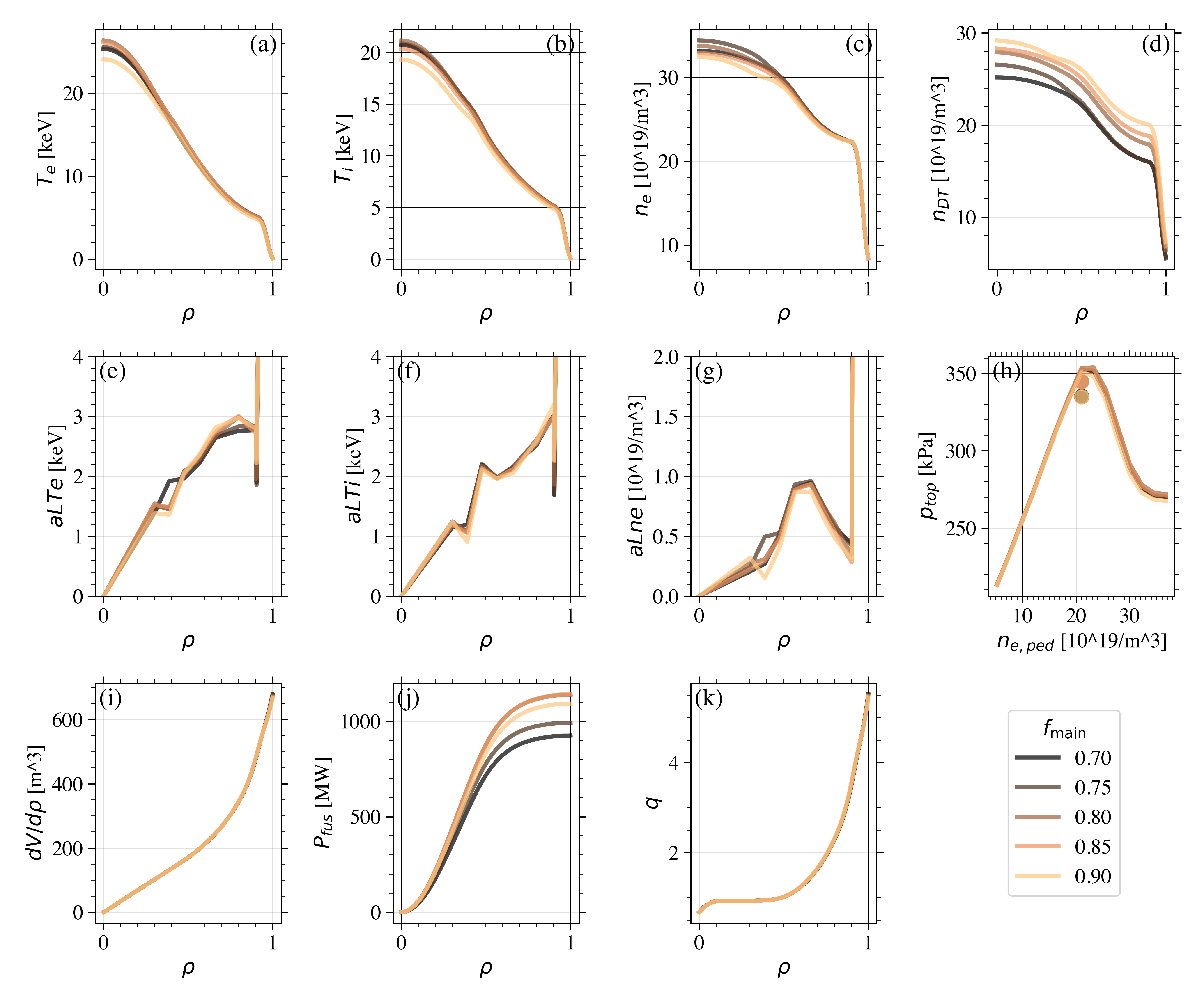}
    \caption{Profiles from scan of $f_{main}$ at nominal value of $Z_{eff}$, $\mathbf{Z_{eff} = 1.5}$, from the 2D scan depicted in Figures \ref{fig:comp_scan} and \ref{fig:comp_scan2}. 
    The electron and ion temperatures are shown in subplots (a) and (b). The electron and deuterium density are shown in subplots (c) and (d). Reduced fuel ion pedestal density due to increasing $f_{main}$, even though $n_{e,ped}$ is constant, can be seen in (d). The inverse normalized electron temperature, ion temperature, and electron density gradient scale lengths are shown in subplots (e), (f), and (g). Minimal ITG stabilization is observed in (f). The sensitivity of pedestal top pressure to $n_{e,ped}$ as calculated by an \texttt{EPED-NN} is shown by the lines in subplot (h), while the full \texttt{EPED} are depicted as points. The \texttt{EPED} dependence on $f_{main}$ is through its impact on $\beta_N$.  At fixed geometry, there is no change in rate of change of the volume with respect to $\rho$ is shown in subplot (i). Subplot (j) shows the total integrated fusion power within a value of $\rho$; fusion power decreases as $f_{main}$ decreases. Finally subplot (k) shows the safety factor profile, which remains unchanged at fixed shaping.}
    \label{fig:profiles_scan_Zeff_1.5}
\end{figure}

\begin{figure}
    \centering
    \includegraphics[width=\linewidth]{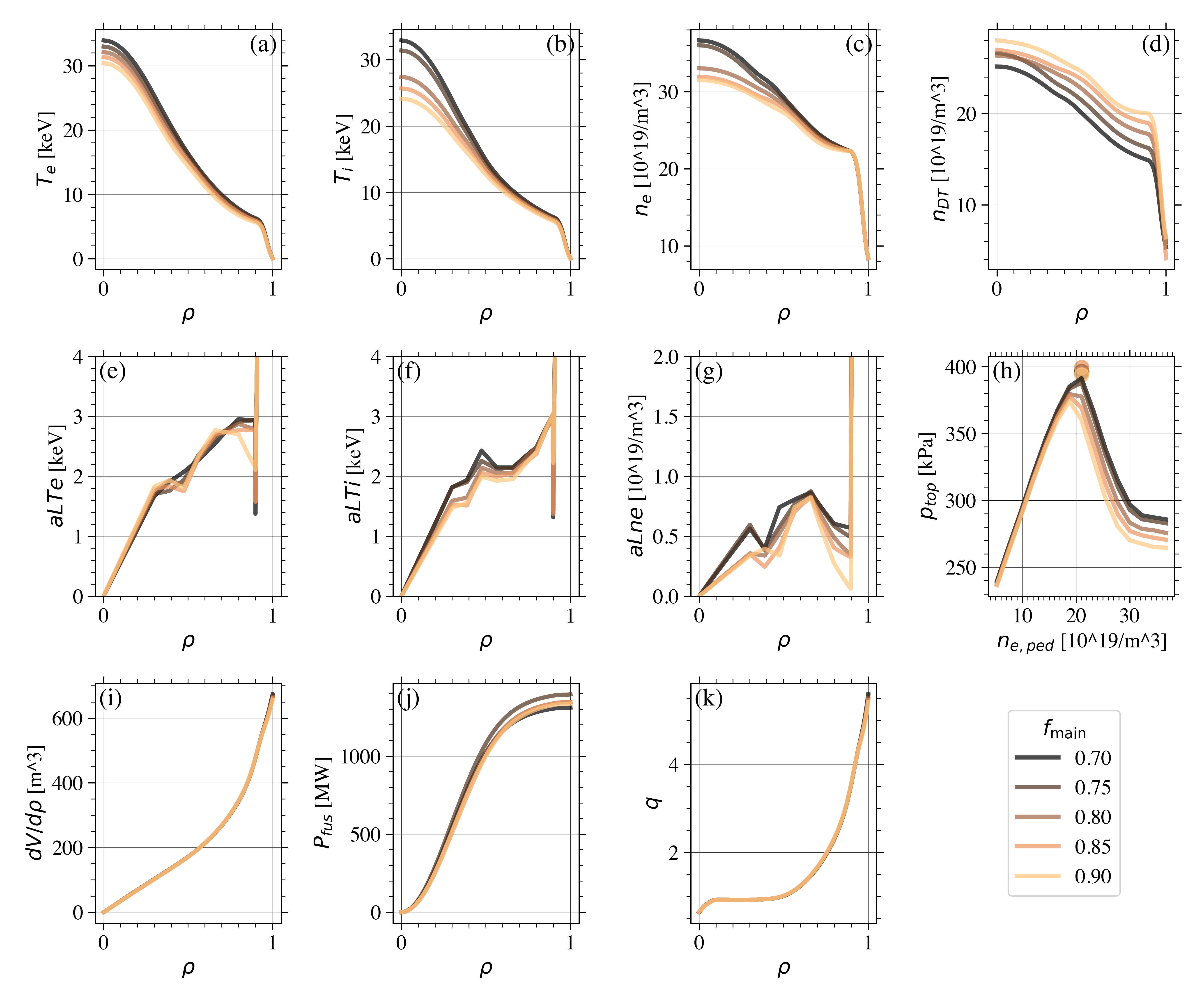}
    \caption{Profiles of $f_{main}$ at elevated $Z_{eff}$, $\mathbf{Z_{eff} = 2.2}$, from the 2D scan depicted in Figures \ref{fig:comp_scan} and \ref{fig:comp_scan2}. 
    The electron and ion temperatures are shown in subplots (a) and (b). The electron and deuterium density are shown in subplots (c) and (d).  Again, there is a reduction in the fuel ion pedestal density duet o increasing $f_{main}$, even though $n_{e,ped}$ is constant. The inverse normalized electron temperature, ion temperature, and electron density gradient scale lengths are shown in subplots (e), (f), and (g). Notable ITG stabilization with decreased $f_{main}$ is observed in (f), and the density gradients increase with $f_{main}$ in (g) due to the change in collisionality. The sensitivity of pedestal top pressure to $n_{e,ped}$ as calculated by an \texttt{EPED-NN} is shown by the lines in subplot (h), while the full \texttt{EPED} are depicted as points. At fixed geometry, there is no change in rate of change of the volume with respect to $\rho$ is shown in subplot (i). Subplot (j) shows the total integrated fusion power within a value of $\rho$. Because the effects of the reduced fuel density and elevated core temperature nearly cancel out, the fusion power remains almost constant with changing $f_{main}$ at this eleveated value of $Z_{eff}$.  Finally subplot (k) shows the safety factor profile, which remains unchanged at fixed shaping.}
    \label{fig:profile_scan_Zeff_2.2}
\end{figure}

\begin{figure}
    \centering
    \includegraphics[width=\linewidth]{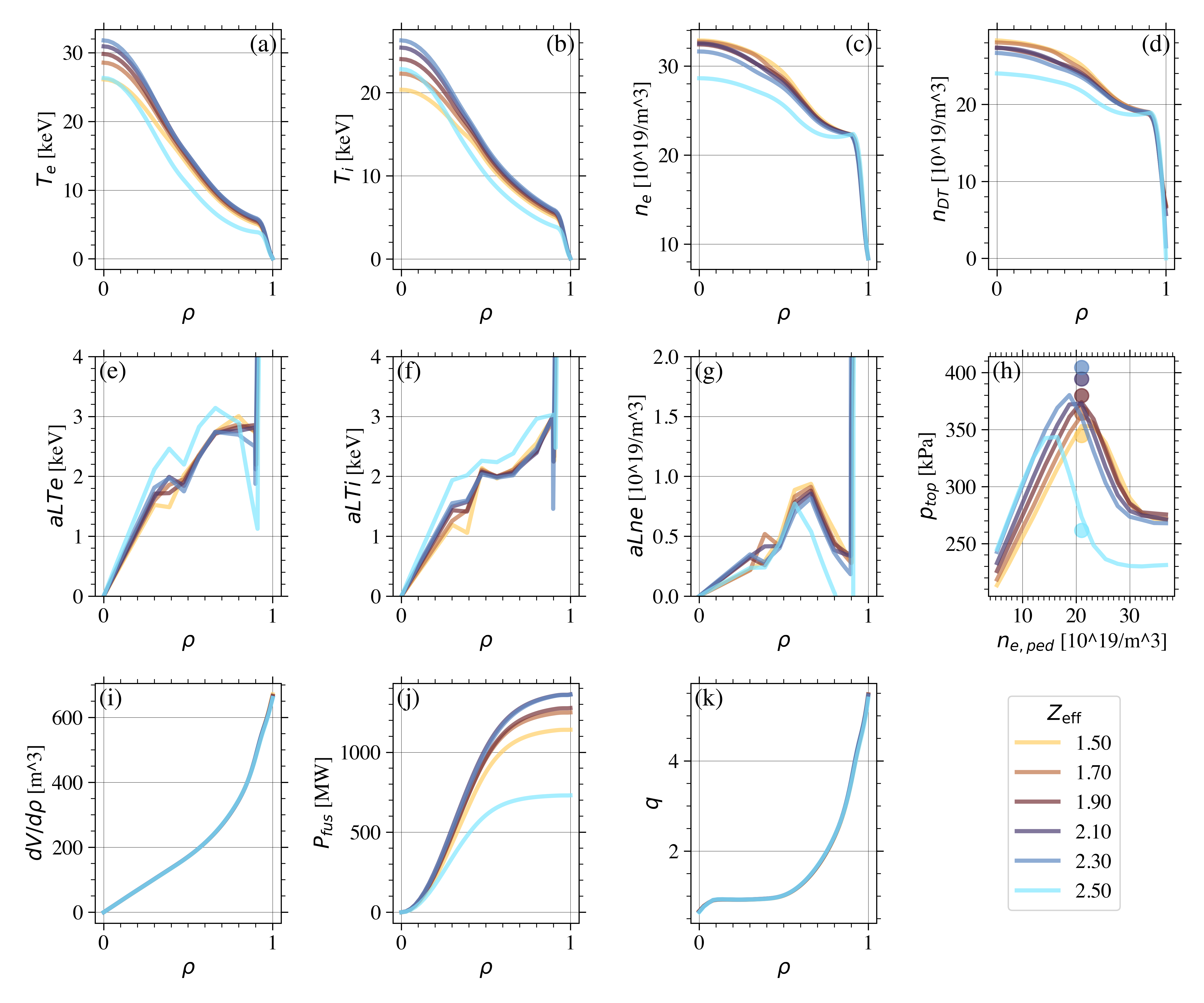}
    \caption{Profiles of $Z_{eff}$ scanned at the nominal value of $f_{main}$, $f_{main} = 0.85 $, from the 2D scan depicted in Figures \ref{fig:comp_scan} and \ref{fig:comp_scan2}. 
    The electron and ion temperatures are shown in subplots (a) and (b). The electron and deuterium density are shown in subplots (c) and (d). The inverse normalized electron temperature, ion temperature, and electron density gradient scale lengths are shown in subplots (e), (f), and (g). Some amount of ITG stabilization is seen in (f). The sensitivity of pedestal top pressure to $n_{e,ped}$ as calculated by an \texttt{EPED-NN} is shown by the lines in subplot (h), while the full \texttt{EPED} are depicted as points. The pedestal pressure, and therefore pedestal temperature, increases with $Z_{eff}$ until the ballooning transition is crossed at $Z_{eff} = 2.5$. At fixed geometry, there is no change in rate of change of the volume with respect to $\rho$ is shown in subplot (i). Subplot (j) shows the total integrated fusion power within a value of $\rho$. Finally subplot (k) shows the safety factor profile, which remains unchanged at fixed shaping.}
    \label{fig:profile_scan_fmain_0.85}
\end{figure}


\begin{figure}
    \centering
    \includegraphics[width=\linewidth]{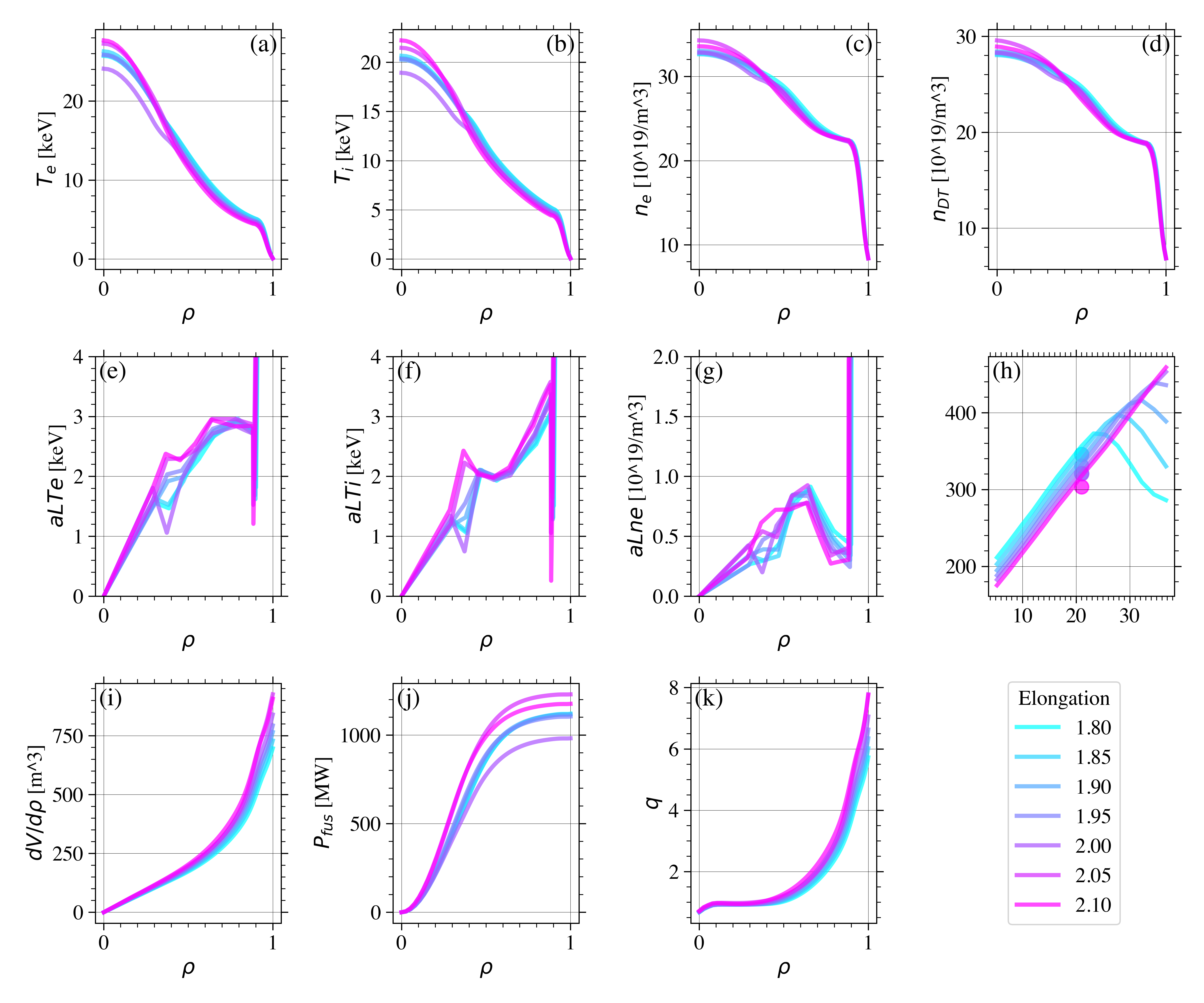}
    \caption{
    Scan of elongation at $\delta = 0.50$ from the 2D scan depicted in Figures \ref{fig:shaping_scan1} and \ref{fig:shaping_scan2}. 
    The electron and ion temperatures are shown in subplots (a) and (b). The electron and deuterium density are shown in subplots (c) and (d). The inverse normalized electron temperature, ion temperature, and electron density gradient scale lengths are shown in subplots (e), (f), and (g). ITG stabilization is observed in (f). The sensitivity of pedestal top pressure to $n_{e,ped}$ as calculated by an \texttt{EPED-NN} is shown by the lines in subplot (h), while the full \texttt{EPED} are depicted as points. The pedestal pressure decreases with increased $\kappa$. Volume increases with $\kappa$, as depicted in subplot (i). Subplot (j) shows the total integrated fusion power within a value of $\rho$, which displays a non-monotonic trend. Finally subplot (k) shows the safety factor profile.}
    \label{fig:kappa_scan_profiles}
\end{figure}

\begin{figure}
    \centering
    \includegraphics[width=\linewidth]{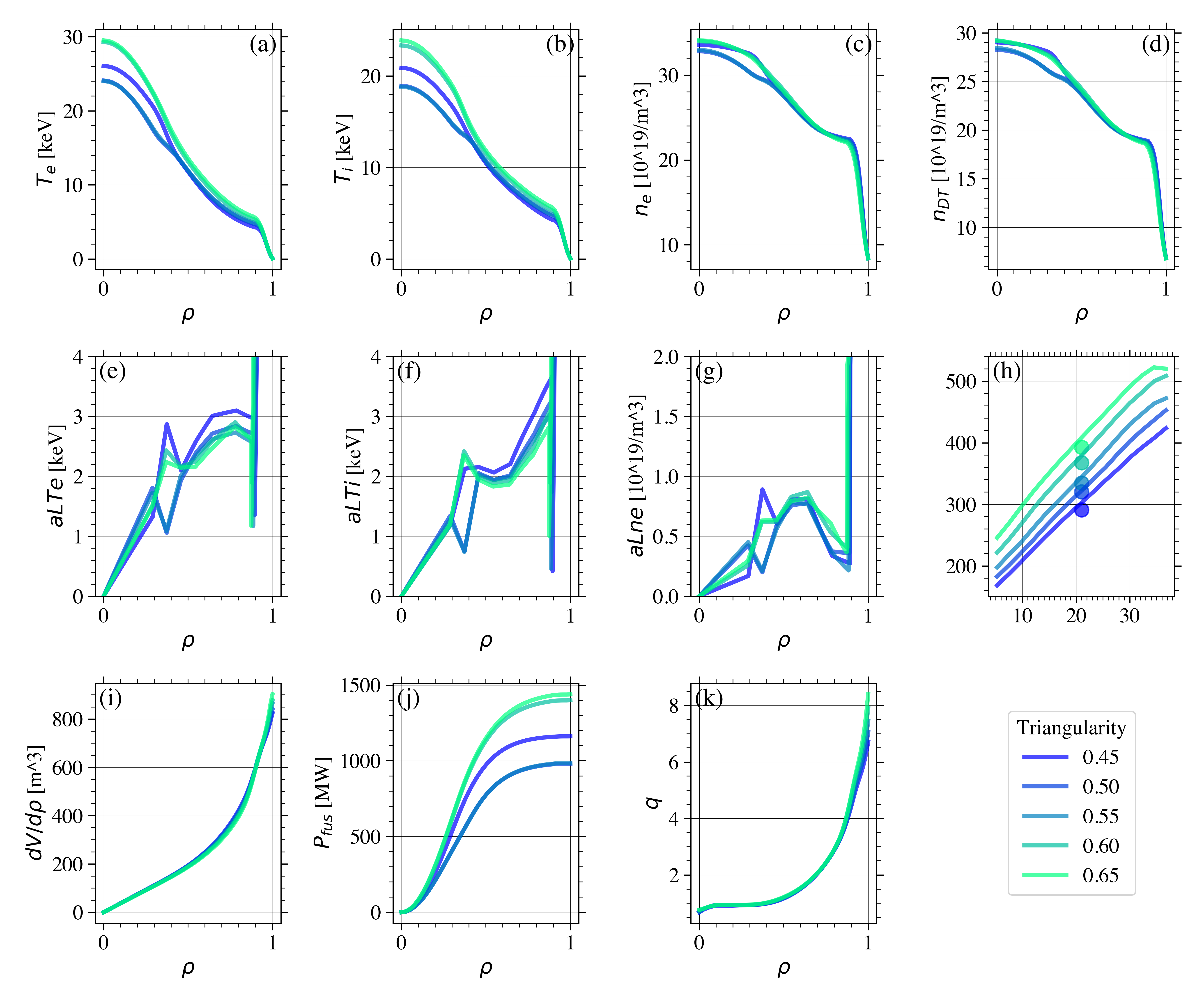}
    \caption{Scan of triangularity at $\kappa = 2.0$ from the 2D scan depicted in Figures \ref{fig:shaping_scan1} and \ref{fig:shaping_scan2}. 
    The electron and ion temperatures are shown in subplots (a) and (b). The electron and deuterium density are shown in subplots (c) and (d). The inverse normalized electron temperature, ion temperature, and electron density gradient scale lengths are shown in subplots (e), (f), and (g). The sensitivity of pedestal top pressure to $n_{e,ped}$ as calculated by an \texttt{EPED-NN} is shown by the lines in subplot (h), while the full \texttt{EPED} are depicted as points. Volume decreases with $\delta$, as depicted in subplot (i).  Subplot (j) shows the total integrated fusion power within a value of $\rho$, which increases with $\delta$. Finally subplot (k) shows the safety factor profile.}
    \label{fig:delta_scan_profiles}    
\end{figure}


\printbibliography
\end{document}